\documentclass[9pt]{elife}

\usepackage{textgreek}

\title{Conformational ensembles of intrinsically disordered proteins and flexible multidomain proteins}

\author[1]{F. Emil Thomasen}
\author[1*]{Kresten Lindorff-Larsen}
\affil[1]{Linderstr{\o}m-Lang Centre for Protein Science, Department of Biology, University of Copenhagen, DK-2200 Copenhagen N, Denmark}

\corr{lindorff@bio.ku.dk}{KLL}

\begin{document}

\maketitle

\begin{abstract}

Intrinsically disordered proteins (IDPs) and multidomain proteins with flexible linkers show a high level of structural heterogeneity and are best described by ensembles consisting of multiple conformations with associated thermodynamic weights. Determining conformational ensembles usually involves integration of biophysical experiments and computational models. In this review, we discuss current approaches to determining conformational ensembles of IDPs and multidomain proteins, including the choice of biophysical experiments, computational models used to sample protein conformations, models to calculate experimental observables from protein structure, and methods to refine ensembles against experimental data. We also provide examples of recent applications of integrative conformational ensemble determination to study IDPs and multidomain proteins and suggest future directions for research in the field.


\end{abstract}

\section{Introduction}

Understanding how proteins carry out their biological functions and what causes them to misfunction is important from the perspective of fundamental science and to develop new therapeutics and biotechnology. Protein dynamics and function are intimately related, and many proteins must modulate their shape to respond to environmental changes, accommodate binding partners, catalyze reactions, convey allosteric signals, and transport ligands \citep{Teilum2009}. Therefore, an important aim of structural biology is not only to determine static structures of proteins, but to characterize their conformational heterogeneity and its relationship to biological function.

Here we review the approaches used to determine conformational ensembles of intrinsically disordered proteins (IDPs) and flexible multidomain proteins, limiting ourselves to approaches that directly integrate experimental data in the generation of the ensemble. The focus will be on principles and examples, and this review is therefore best suited for readers who are looking for a brief, conceptual overview of the field. For a more technical introduction we refer the reader to previous literature \citep{Hummer2015,Orioli2020}.

\subsection{Intrinsically disordered proteins and multidomain proteins}

IDPs are proteins that do not fold into a well-defined structure, but rather interconvert between a large range of very different conformations. Based on disorder predictions, intrinsically disordered regions (IDRs) make up more than one third of eukaryotic proteins \citep{Ward2004,Xue2012,Oates2013,Tunyasuvunakool2021}. The conformational heterogeneity of IDPs allows for promiscuity in interaction partners, and IDPs are often involved in biological processes such as signaling, recognition, and regulation \citep{Wright2015,Bondos2021}. In recent years, IDPs have received much attention for their role in the formation of a number of biomolecular condensates and membraneless organelles \citep{Banani2017,Boeynaems2018,Peran2020,Dignon2020,Choi2020,Martin2020,Borcherds2021}. Condensate formation can be driven by interactions between a variety of different biomolecules, including IDPs, folded proteins, and RNA, and understanding the molecular details that underly condensate formation is an important motivation for structural characterization of IDPs and their interactions.


Many proteins are modular, consisting of folded domains connected by flexible linkers or IDRs \citep{Apic2001,Ekman2005}. Such multidomain proteins can display a high level of conformational heterogeneity, as the folded domains can rearrange with respect to each other \citep{Wriggers2005,Roy2016,10.3389/fmolb.2016.00054}. For this reason, many of the approaches used for structural characterization of IDPs also extend to flexible multidomain proteins. Here, we will use the term multidomain protein to refer to structurally dynamic multidomain proteins with flexible linkers.

Due to their structural heterogeneity, IDPs and multidomain proteins are best described by an ensemble of structures representing the conformations sampled by the system with associated thermodynamic weights. In this review, we will focus on the integrative approaches used to determine conformational ensembles of IDPs and multidomain proteins. Further, we will discuss applications of conformational ensemble determination using recent examples from the literature, and will suggest future directions for research in the field.


\begin{figure}[hbt!]
\begin{fullwidth}
\includegraphics[width=0.95\linewidth]{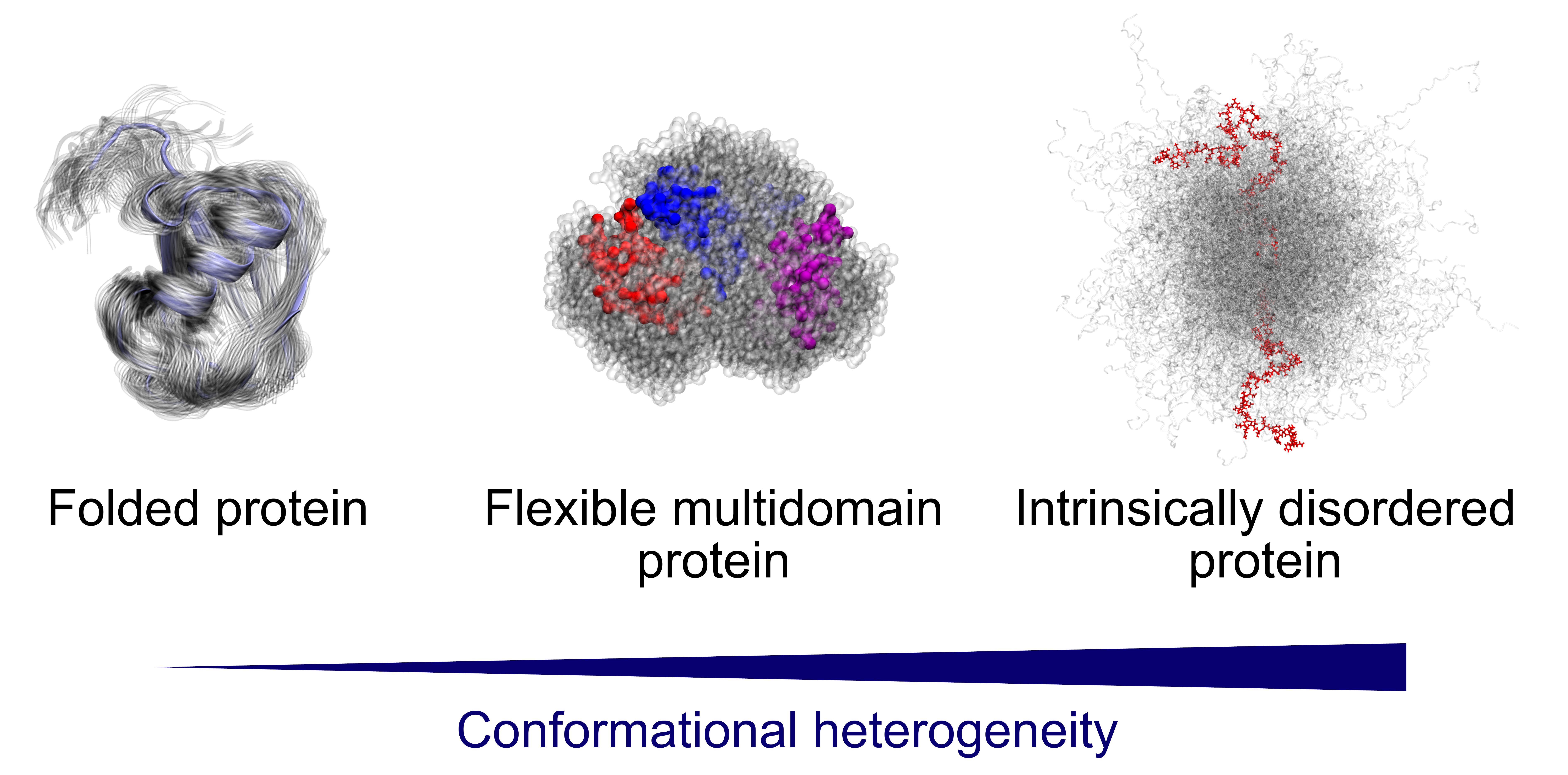}
\caption{\textbf{Conformational heterogeneity in proteins}\\Proteins do not exist as rigid structures in solution, but rather sample an ensemble of structures. Different proteins have variable levels of conformational heterogeneity. Stably folded proteins undergo relatively small conformational fluctuations around an average structure (left). Multidomain proteins consisting of folded domains connected by flexible linkers can display a higher level of conformational heterogeneity, as the folded domains can rearrange with respect to each other (middle). Intrinsically disordered proteins are characterized by a high level of conformational heterogeneity, as they do not fold into a well-defined structure, but rather interconvert between a range of conformations (right). The examples shown here are ubiquitin (folded protein) \citep{Lindorff-Larsen2005}, TIA-1 (flexible multidomain protein) \citep{Larsen2020,Thomasen2021}, and $\alpha$-synuclein (intrinsically disordered protein) \citep{Thomasen2021}.}
\label{fig:1}
\end{fullwidth}
\end{figure}


\section{How do we determine conformational ensembles of IDPs and multidomain proteins?}

Most biophysical experiments used to study protein structure report on observables that are averaged over different conformations. Determining atomic resolution conformational ensembles of proteins from sparse, ensemble-averaged experimental data without further constraints is an ill-posed problem, as it involves fitting many more distributions of atomic positions than there are data-points available. Computational models and simulations can be used to sample conformational ensembles of proteins at atomic resolution. However, computational models have various limitations, especially for modeling IDPs, and are not always in agreement with experiments \citep{Palazzesi2015,Rauscher2015,Henriques2015,Bottaro2018,Robustelli2018}. Thus, neither biophysical experiments nor computational modeling alone provide the optimal solution for conformational ensemble determination. However, combining both using integrative methods can provide conformational ensembles that are also in agreement with experimental observations. This process usually involves four central components: (1) a computational model for sampling protein conformations, (2) one or more biophysical experiments that report on protein structure, (3) corresponding models for calculating experimental observables from the conformational ensemble (so called forward models), and (4) a method for refining the ensemble based on experimental data. Depending on the approach, experimental data can either be integrated directly in the conformational sampling procedure or after sampling.

\subsection{Experimental methods}

For a biophysical technique to be useful in integrative ensemble determination of IDPs and multidomain proteins, it should report on protein structure under solution conditions where conformational heterogeneity is present. We write `report on' because solution conditions are not a strict requirement; methods that use sample freezing, such as electron paramagnetic resonance (EPR), cryo-electron microscopy (cryo-EM and solid-state NMR), may capture the conformations present in solution if cooling is rapid \citep{Fischer2010,hu2010detection,Chen2019,Bock2021,Klose2021}. Additionally, for an experiment to be useful, an accurate forward model must be available to calculate the experimental observable from the ensemble. 

Generally, the more different types of experiments that are integrated, the more trustworthy the resulting ensemble. Combining experiments that report on different structural features ensures that the ensemble is accurate at multiple structural levels and can prevent overfitting to a single experiment \citep{Gomes2020,naudi2021quantitative}. Another good approach to prevent overfitting is to use subsampling of the experimental data or to leave some data out of ensemble determination for later validation.

Small angle X-ray scattering (SAXS) experiments are often used in conformational ensemble determination. SAXS is a low-resolution technique that reports on the overall shape and size of the protein in solution. For IDPs and multidomain proteins, SAXS contains information on the global dimensions of the ensemble and can for example capture domain rearrangements in multidomain proteins \citep{Yang2010,Rozycki2011,Bernado2012,Tuukkanen2017}. 

Another commonly used technique is nuclear magnetic resonance (NMR) spectroscopy. NMR experiments can report on a variety of structural features. For example, chemical shifts depend on the local environment of nuclei and contain information on backbone conformations and secondary structure \citep{ozenne2012mapping,Kragelj2013}, scalar \emph{J}-couplings report on bond connectivity and dihedral angles, nuclear Overhauser effects (NOEs) report on short-range distances between nuclei, and residual dipolar couplings (RDCs) report on the alignment of bond vectors with respect to a global alignment tensor \citep{Marion2013}. 

By chemically modifying the protein with a paramagnetic spin-label, transient long-range interactions between nuclei and the spin-label modification can be probed by NMR using paramagnetic relaxation enhancement (PRE) experiments \citep{Clore2009}. Similarly, double electron-electron resonance (DEER) experiments probe long-range distances between two paramagnetic spin-labels using EPR \citep{Pannier2000}.

Other methods used for integrative conformational ensemble determination include single molecule Förster resoncance energy transfer (smFRET), which reports on the distance between two sites in the protein chemically labeled with a fluorescence donor and acceptor dye \citep{Metskas2020,Lerner2021,Alston2021}, infrared (IR) spectroscopy, which is mostly used to report on protein secondary structure and local electrostatic environment based on amide I vibrational frequencies \citep{Haris2013,Reppert2013}, and cryo-EM, which can in principle resolve the conformational ensemble at atomic resolution, but where the electron density map can also be considered an ensemble-averaged observable \citep{Cossio2013,Bonomi2018,Bonomi2019}.

\subsection{Conformational sampling}

One must choose a computational method to sample or generate protein conformations. A common approach is to use molecular dynamics (MD) or Monte Carlo (MC) simulations \citep{Hollingsworth2018,Vitalis2009,Braun2019}, where conformations are sampled based on the energy specified by a force field, a potential energy function describing bonded and non-bonded interactions between all atoms. There are two central limitations to such simulations: (1) force fields contain inaccuracies due to inherent approximations and parameterization (2) full sampling, meaning that all conformations have been sampled with the correct weights, can be computationally expensive or infeasible \citep{Bottaro2018}. In recent years, popular force fields for proteins have been modified to improve accuracy for IDPs \citep{Best2014,Piana2015,Huang2017,Robustelli2018,Zerze2019,Thomasen2021}.

Coarse-grained (CG) models, where system complexity is reduced by mapping groups of atoms to single particles, can alleviate problems with sampling in biomolecular simulations, but come at the cost of accuracy and resolution \citep{Ingolfsson2014,Kmiecik2016}. However, because the generated ensembles are optimized to be in accordance with experimental data, CG simulations are a useful tool for integrative ensemble determination. Since most observables are a function of atomic positions, forward models must be developed specifically for the CG model or CG ensembles must be back-mapped to all-atom structures in order to use conventional forward models, partly undermining the gained computational efficiency. A popular CG model for biomolecular systems is Martini \citep{Marrink2007,Monticelli2008,Souza2021}, which has previously been used to simulate multidomain proteins and IDPs with slight modifications to the force field \citep{Berg2018,doi:10.1098/rsfs.2018.0062,Larsen2020,doi:10.1126/sciadv.abc3786,Martin2021,doi:10.1126/sciadv.abh3805,Benayad2021,Thomasen2021}. Additionally, numerous CG models have been developed specifically to study IDPs \citep{Dignon2018,Regy2021,Tesei2021a,Wu2018,Cragnell2016,Cragnell2018,Vitalis2009,Das2018a,Das2018,Choi2019,Mioduszewski2018,Rutter2015}.

There are also less computationally demanding methods to generate conformational ensembles. For example, Flexible-Meccano generates IDP conformations by sampling backbone dihedrals using information from non-secondary structural elements of existing protein structures  \citep{Ozenne2012}. Conformational propensities such as transient secondary structure and long-range interactions can be included based on prior knowledge. In a similar approach to Flexible-Meccano, IDP ensembles can also be constructed from a fragment library sampled by MD simulations \citep{Pietrek2020}. For multidomain proteins, conformations can be generated with Pre\_bunch, a part of the BUNCH program \citep{Petoukhov2005,Bernado2007}. Here conformations are generated by sampling C\textsubscript{\textalpha}-C\textsubscript{\textalpha} dihedrals in linkers, while folded domains are treated as rigid bodies.

\subsection{Calculating experimental observables}

Linking experiments with an underlying conformational ensemble requires forward models to calculate experimental observables from the ensemble. In most cases, the observable is calculated from each static structure of the ensemble and subsequently averaged. Thus, effects of dynamics (i.e. the timescales of interconversion) on the observable are neglected if these are not implicitly included in the forward model, and one must keep in mind that this is an approximation for time-dependent processes such as several types of NMR and fluorescence measurements.

Some experimental observables have accurate forward models available. A good example is the calculation of SAXS intensities, which are a function of all interatomic distances, and are well-described by the Debye scattering equation, although SAXS forward models often use approximations of the Debye equation to increase computational efficiency \citep{Hub2018,Svergun1995,Petoukhov:fs5015,Schneidman-Duhovny2010,Schneidman-Duhovny2013,Gumerov2012,Grudinin2017}. 

Many experiments report on distances, which are straightforward to calculate from ensembles. For example, NOEs are often calculated simply as an $r^{-6}$-weighted ensemble-average of interatomic distances $r$, with the approximation that dynamics do not contribute to the NOE intensity \citep{Brueschweiler1992,Peter2001,Smith2020}. Distance-based experiments that require chemical labeling, such as smFRET, PRE, and DEER, introduce an additional challenge, as label dynamics must be taken into account in the forward model \citep{Steinhoff1996,Tombolato2006a,Tombolato2006,Salmon2010,Polyhach2011,Sindbert2011,Kalinin2012,Reichel2018,Borgia2018,Tesei2021,Klose2021}.

For other experiments, the relationship between structure and observable is less straightforward. For example, the relationship between NMR chemical shifts and the local environment of nuclei is complex, so forward models are usually empirically optimized against data on many proteins \citep{Kohlhoff2009,Shen2010,Han2011}. In some cases, the lack of an accurate forward model is the main limitation in the use of experimental data. For example, circular dichroism (CD) spectra contain useful structural information, but are usually interpreted as a sum of basis spectra determined for folded proteins, making CD difficult to use for IDPs \citep{Nagy2019,jephthah2021force}. When studying IDPs, it is a general problem that empirically-derived forward models are often trained on folded proteins, and may not capture e.g. solvation properties of IDPs \citep{Piana2015,Henriques2018,Pesce2021}. However, it is challenging to obtain training sets of accurate IDP ensembles with experimental data available that are independent of the forward model one wishes to optimize \citep{Lindorff-Larsen2021}.

Forward models often require system-specific setting of parameters and care must be taken to avoid overfitting to experimental data. One example is the calculation of SAXS profiles from protein structures, which often involves implicit modeling of the solvation shell. This presents a dilemma, as fitting solvation shell parameters against experimental data individually for each conformation likely leads to overfitting, but a global fit for all conformations may not capture conformation-dependent changes in the hydration shell \citep{Pesce2021}. Another example is the calculation of RDCs, where fitting the alignment tensor against experimental data individually for each frame may result in overfitting and there may be parameter correlation between the axial component of the alignment tensor and the order parameter that describes the dynamics of the internuclear vector \citep{Zweckstetter2008}.


\begin{figure}[hbt!]
\begin{fullwidth}
\includegraphics[width=0.95\linewidth]{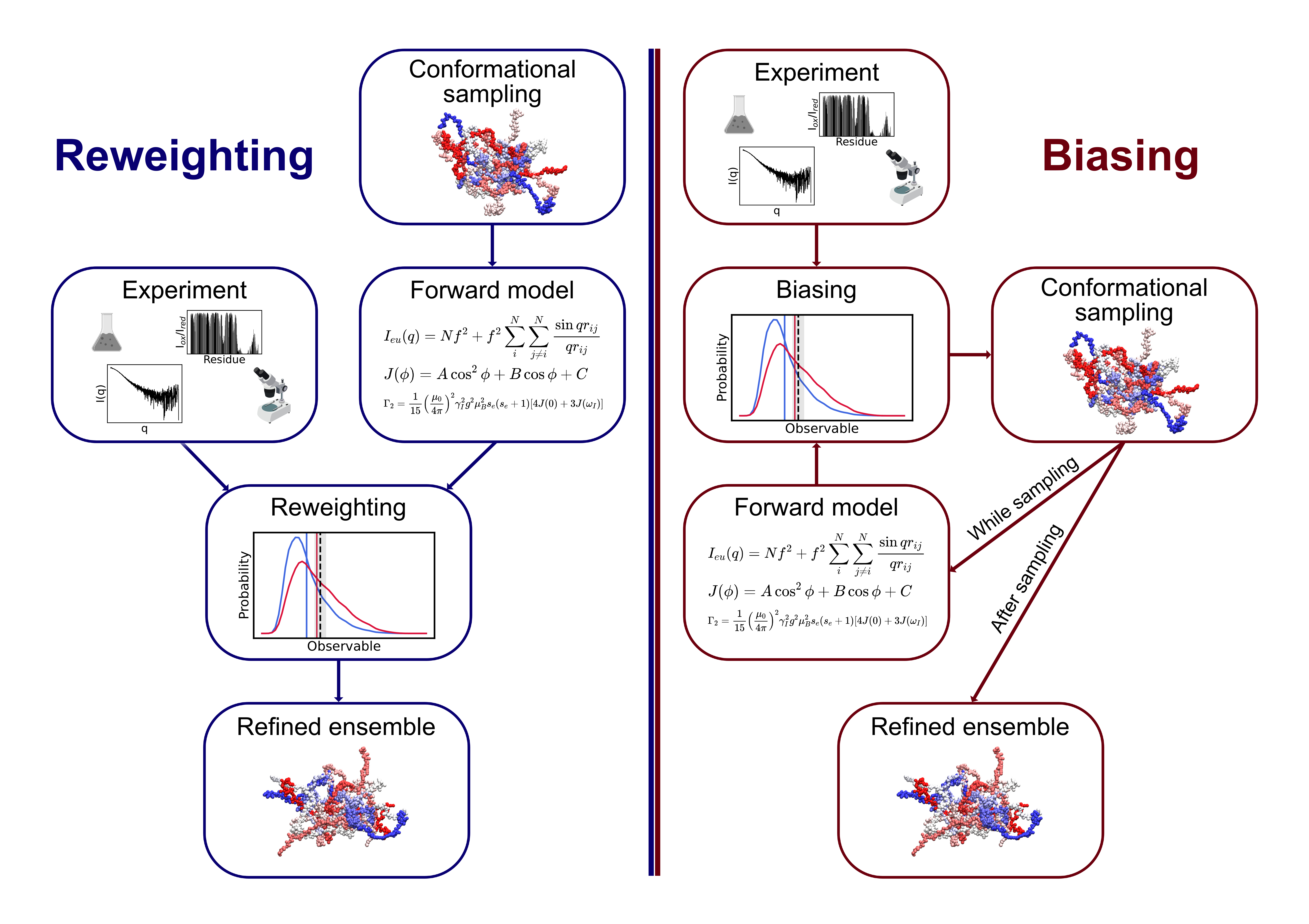}
\caption{\textbf{Integration of experimental data and computational model by reweighting or biasing}\\Reweighting and biasing are the two central approaches used to integrate experimental data and computational models in conformational ensemble determination. In reweighting, unbiased conformational sampling is performed initially. Subsequently, experimental observables are calculated from the ensemble using forward models, and conformations are reweighted to improve agreement with experimental data. The optimization of the weights is usually regularized by a maximum-entropy constraint to minimally perturb the initial ensemble. In biasing, experimental data is integrated directly in the conformational sampling by biasing the force field. Experimental observables are calculated iteratively while running the simulation, and agreement with experimental data is imposed either through simple restraints over multiple simulation replicas or by on-the-fly modification of the force field, usually in accordance with the principle of maximum-entropy.}
\label{fig:2}
\end{fullwidth}
\end{figure}


\subsection{Integrating the modeled ensemble and experimental data}

Integrative ensemble determination is usually based on refining the modeled ensemble to improve agreement with experimental data. This can either be done by biasing the conformational sampling procedure or by reweighting the probability distribution of conformations after sampling. 

Two distinct approaches are commonly used to modify the initial ensemble based on experimental data; maximum-entropy (MaxEnt) and related Bayesian methods minimally perturb the initial distribution of conformations to optimize the agreement with experimental data \citep{PhysRev.106.620}, whereas maximum-parsimony (MaxPars) methods select the fewest number of conformations required for agreement with experimental data. As ensembles of IDPs and multidomain proteins are not expected to be captured well by few conformations, MaxEnt methods are most suitable for these cases.

In MaxEnt approaches, the minimal perturbation of the initial ensemble is achieved by optimizing the relative Shannon entropy (or negative Kullback-Leibler divergence) between the initial ensemble and the refined ensemble \citep{Shannon1948,Kullback1951}. In MaxEnt reweighting, the MaxEnt regularization is included as a constraint in the optimization of the ensemble weights \textit{after} generating the initial ensemble \citep{Rozycki2011}. In MaxEnt biasing methods, sampling and MaxEnt ensemble refinement are performed simultaneously by modifying the forces on the fly to optimize agreement between ensemble and experiment. Alternatively, MaxEnt biasing can be achieved by simulating multiple replicas and restraining the average over the replicas to agree with experiment \citep{Pitera2012,Roux2013,Cavalli2013,Boomsma2014}.

Reweighting and biasing methods are commonly expressed in the framework of Bayesian statistics to explicitly include sources of error and noise. Both MaxEnt and MaxPars approaches can be expressed in Bayesian formalisms \citep{Fischer2010,Olsson2013,Hummer2015}, but the Bayesian frameworks also opens up for the possibility to use other priors and likelihood functions than those given by MaxEnt and MaxPars \citep{Orioli2020}. A plethora of different reweighting and biasing methods exist, applying different combinations of MaxEnt, MaxPars, and Bayesian statistics. For an extensive list, see \cite{Bonomi2017}.

Should one choose reweighting or biasing methods for ensemble optimization? An advantage of the reweighting approach is that new data can be included to improve the ensemble at any time after generating the initial ensemble. Additionally, one can use initial ensembles from other sources than molecular simulations, such as Flexible-Meccano \citep{Ozenne2012}. A disadvantage of the reweighting approach is that all structures in the refined ensemble must already be present in the initial ensemble. Therefore, reweighting is only reliable when the initial ensemble is reasonably accurate, and the biasing approach is preferable if the force field is known to be inaccurate or the sampling is poor. However, biasing approaches for MD sampling require that forward models are differentiable and that the forward model calculations are fast as they are executed at (almost) every integration step in the simulations. Finally, it is possible to combine experimentally-biased sampling and post-simulation integration of experimental data \citep{shen2008consistent,boomsma2014equilibrium,Hummer2015,rangan2018determination,stelzl2021global}

\section{What can conformational ensembles tell us?}

Integrative conformational ensemble determination can be viewed as a way of combining our general knowledge about the chemistry and physics of proteins, encoded for example in the force field or other conformational sampling model, with our interpretation of the experimental data. Thus, conformational ensembles can be used to examine detailed properties of the system that are not directly available from the experimental data, but which are likely accurate given the data and our prior knowledge about proteins. In some cases, conformational ensemble determination can provide a consistent structural interpretation of seemingly inconsistent experimental results, illustrating the strength of the approach \citep{Fuertes2017,Gomes2020}.

While much of the research in integrative conformational ensemble determination has been focused on developing new methods and proof-of-concept, the field has advanced to a point where it can address biological questions from the ensemble perspective. Here, we provide some recent examples where conformational ensembles have been applied to study IDPs and multidomain proteins.

\subsection{Multidomain proteins}

Conformational ensembles have been used to study flexibility and domain interactions in multidomain proteins. For example, \cite{Weber2018} used MD simulations biased using RDC data from NMR using metadynamics metainference \citep{Bonomi2016}, to generate conformational ensembles of the antibody light chain, a multidomain protein with two folded domains connected by a flexible linker. They showed that a mutation in the linker increases flexibility and decreases interdomain contacts, illustrating the importance of the linker sequence in maintaining correct domain orientations and reducing amyloidogenicity.

In another example, \cite{Martin2021} determined conformational ensembles of the multidomain protein hnRNPA1 using the CG MD model Martini \citep{Marrink2007,Monticelli2008}, SAXS and a Bayesian MaxEnt (BME) reweighting approach \citep{Bottaro2020} to study interactions between folded domains and an IDR as a function of ionic strength. Based on the correlation between folded domain-IDR contacts and phase separation propensity, they proposed a model for how the folded domains of hnRNPA1 modulate its ability to form biomolecular condensates.

\cite{doi:10.1126/sciadv.abc3786} used Martini, SAXS, and metainference to study linear polyubiquitin and its binding to NEMO, a regulator of inflammation, which preferentially binds to long polyubiquitin chains. They determined conformationa ensembles of two, three, and four-ubiquitin chains in their NEMO-bound and unbound states. Their results suggest that linear polyubiqutin behaves as a self-avoiding polymer, that NEMO binding greatly restricts conformational flexibility, and that the likelihood of finding a NEMO binding site in a bound-like conformation increases with chain length, in line with the preferential binding to longer chains.

\subsection{Multidomain protein-nucleic acid complexes}

Integrative conformational ensemble determination is not limited to isolated multidomain proteins; the approach has recently been applied to study complexes between multidomain proteins and nucleic acids. \cite{Kooshapur2018} used SAXS data and metainference MD simulations to study the complex between the RNA-binding domains of hnRNPA1 and the micro-RNA pri-mir-18a, which is processed by hnRNPA1. The conformational ensemble was consistent with recognition of two UAG-motifs in the terminal loop of pri-mir-18a and revealed that the upper stem of pri-mir-18a is partially melted in the complex.

In another example, \cite{Saad2021} used SAXS data and metainference MD simulations to investigate the heterodimeric transcription factor E2F1/DP1 bound to DNA. Their results revealed that the complex is highly dynamic and that flexibility in DP1 may be important for stabilizing interactions with DNA, contextualizing cancer-related mutations found in flexible regions of the protein.

\subsection{IDRs and membranes}

Recently, integrative conformational ensemble determination has been used to study IDRs in the context of lipid membranes. \cite{Pond2020} integrated neutron reflectometry data and MD simulations using MaxEnt biasing to study interactions between a lipid-bilayer and the intrinsically disordered SH4-U domain of Hck kinase, which anchors Hck kinase to the membrane. Their results suggest that SH4-U preferentially interacts with anionic lipids and can insert itself into the lipid-bilayer through interactions between charged residues and lipid head-groups.

In another example, \cite{doi:10.1126/sciadv.abh3805} used Martini, SAXS, and BME reweighting to investigate the conformational ensemble of the human growth hormone recpetor (hGHR) in a nanodisc. hGHR is a single-pass transmembrane protein with a folded extracellular domain and a disordered intracellular domain. The conformational ensemble of hGHR revealed that the intracellular domain does not interact strongly with the membrane, but instead extends into the cytosol, giving it a large capture radius for interaction with its many binding partners.

\subsection{Multimodal conformational ensembles}

Conformational ensembles are especially useful for investigating systems that populate several distinct populations and systems with interesting subpopulations, as information on these properties are inherently not contained in ensemble-averaged observables. For example, \cite{Reppert2016} determined conformational ensembles of fragments from the IDP Elastin using MD simulations and MaxEnt reweighting against amide I IR spectroscopic data. These fragments each consisted of different repeated sequence-motifs thought to be important for the elastic properties of Elastin. For all different motifs, the ensembles revealed a bimodal distribution of conformations centered around a collapsed kinked state and an extended state, with the population of each state determined by small differences in the sequence of the motifs. The authors proposed that the elastic properties of Elastin are related to the bimodality of these motifs and that the motif-composition of Elastin is evolutionarily tuned to maintain disorder and elastic properties.

In another example, \cite{Huang2020} investigated the structural effect of linker mutations that increase the kinase activity of the multidomain protein Hck kinase. Using as a starting point nine conformations of Hck kinase previously determined with CG simulations \citep{Yang2010}, they applied the reweighting method BSS-SAXS, which optimizes the weights of a small conformational ensemble without MaxEnt regularization or MaxPars selection \citep{Yang2010}, to reweight the conformations against SAXS data on both the wild-type and mutant protein. This revealed a population shift from the "assembled", inhibited state of the protein to the "disassembled", active state of the protein, demonstrating the importance of linker regions in tuning kinase activity.

\subsection{Small-molecule binding to IDPs}

It has recently been demonstrated that integrative ensemble determination can be used to study small-molecule binding to IDPs. The important role of IDPs in signalling and regulation and their propensity to form disease-related aggregates makes them interesting targets for drug development \citep{Metallo2010,Dunker2010,Heller2015,Heller2018}. \cite{Heller2017} investigated the binding of a small-molecule to the IDP c-Myc using metadynamic metainference MD simulations \citep{Bonomi2016} biased against NMR chemical shifts. They found that the small-molecule does not bind to a specific site on c-Myc, but rather diffuses across the protein with increased affinity for certain residues, which were in agreement with experimental mutational studies.

In a related study, \cite{doi:10.1126/sciadv.abb5924} investigated the binding of a small-molecule shown to inhibit amyloid aggregation to amyloid-$\beta$. Again they used metadynamic metainference simulations with NMR chemical shifts. Their results showed that small-molecule binding increases the conformational heterogeneity of amyloid-$\beta$, suggesting that binding may be driven by entropic expansion of the IDP. In these two studies, the ensemble perspective shed light on the mechanisms underlying small-molecule binding to IDPs, and the results suggest general principles for designing drugs to target IDPs.

\section{Future perspectives}

In this section, we will highlight five areas of research which we think will be important for the future of integrative conformational ensemble determination.

First, the development of accurate and transferable force fields is an important element in advancing integrative ensemble determination. We note that force fields are generally improving \citep{Lindorff-Larsen2012,Beauchamp2012}, and as mentioned previously, specific attention has been given to improving force fields for IDPs \citep{Best2014,Piana2015,Huang2017,Robustelli2018,Zerze2019,Thomasen2021}, and importantly developing force fields that work well for both folded proteins and IDPs \citep{Robustelli2018}. Applications of machine learning, Bayesian inference, maximum likelihood and reweighting approaches to automate force field optimization based on quantum mechanical calculations and biophysical experiments show potential for the development of more accurate force fields \citep{Norgaard2008,Li2010,Wang2014,Kofinger2021,Tesei2021a,Noe2020,Gkeka2020,Lindorff-Larsen2021,Unke2021,Yang2021.04.26.441401}.

Second, an important area of future research is the development of forward models to calculate observables from structure. Focus should be given to developing forward models that are transferable between folded proteins and IDPs. One of the challenges lies in choosing which data to train such forward models on. Ideally, models should be trained on IDP ensembles that are already well determined, but this quickly becomes a "chicken or the egg"-type problem. One possibility may be to optimize the forward model and IDP conformational ensemble in a self-consistent manner using reweighting approaches.

Third, a future prospect is to exploit recent advances in protein structure prediction. All of the conformational sampling methods discussed in this review require that the structures of the folded domains are known initially. Previously, this meant that an experimentally determined structure of each folded domain in the protein must be available. Substantial improvements in the accuracy of protein structure prediction\citep{Jumper2021} open up for the possibility to perform high-throughput ensemble determination of multidomain proteins with experimental data available using for example CG simulations.

Fourth, a potential development in the field is to determine conformational ensembles based on time-resolved experiments or time-dependent data that depend on the time-scales on which motions occur. The experiments discussed here all report on static observables that are simply an average over all conformations of the ensemble. However, for certain types of biophysical experiments, time-dependent dynamics must be taken into account. For example, NMR relaxation experiments report on molecular motions which can only be calculated from a time-series of structures or models of the dynamics, such as the Lipari-Szabo model-free approach \citep{Lipari1982,Brueschweiler1992,Peter2001,Salvi2016,Smith2020,Kummerer2021}. Additionally, many biological processes do not happen at equilibrium, but rather involve transitions from one state to another. Such transitions can be measured by many of the biophysical techniques mentioned in this review. For example, time-resolved SAXS and NMR experiments can be used to measure kinetic processes such as conformational changes and binding \citep{Rennella2013,Tuukkanen2017,Cho2021}. While approaches have already been developed to interpret NMR relaxation data using conformational ensembles \citep{Salvi2016,Kummerer2021}, future research could involve the development of methods exploiting time-resolved data to resolve biologically relevant transitions.

Fifth, in order to move to more complex systems we need to be able to study conformational ensembles of heterogeneous or polydisperse mixtures of proteins. The integrative modeling approaches discussed in this review are based on calculating averaged observables from ensembles of isolated protein species and refining ensembles against experimental data measured on conformationally heterogenous but chemically homogeneous samples of pure proteins. Future research in the field could involve the development of methods that use experimental data not only averaged over conformations but also over a mixture of different molecular species. This would be useful in cases where specific species cannot be isolated, such as for polydisperse oligomers. Such methods would involve the extraction of \textit{even more} information from sparse experimental data, so care will have to be taken to prevent overfitting of ensembles.

\section{Conclusion}

In this review, we have discussed the steps that go into determining conformational ensembles of IDPs and multidomain proteins using integrative approaches. Choosing the right combination of experiments, computational modeling, forward models, and integration approach will ensure that the resulting ensemble is accurate with regards to different structural features. We have also discussed recent applications of integrative conformational ensemble determination, showing that the field is ripe for addressing real biological questions from the ensemble perspective. Finally, we have proposed future directions for the field, including the development of more accurate and transferable force fields and better forward models for IDPs. We have also discussed the potential of interpreting time-resolved experiments and heterogenous protein samples using conformational ensembles. These advances can bring the field beyond static thermodynamic ensembles of simple systems and towards modeling of dynamic biological processes and complex polydisperse mixtures.

\section{Acknowledgments}
We acknowledge numerous fruitful discussions with members of the Lindorff-Larsen group on integrative modelling. Our work in this area is supported by the Lundbeck Foundation BRAINSTRUC initiative (R155-2015-2666 to KL-L).

\bibliography{references}

\begin{thebibliography}{159}
\providecommand{\natexlab}[1]{#1}
\providecommand{\urlprefix}{}
\providecommand{\doiprefix}{doi: }

\bibitem[{Alston et~al.(2021)Alston, Jhullian J and Soranno, Andrea and
  Holehouse, Alex S}]{Alston2021}
\textbf{\color{eLifeMediumGrey} Alston JJ}, Soranno A, Holehouse AS.
\newblock {Integrating single-molecule spectroscopy and simulations for the
  study of intrinsically disordered proteins}.
\newblock Methods.  2021; 193:116--135.
\newblock
  \urlprefix\url{https://www.sciencedirect.com/science/article/pii/S104620232100092X},
  \href{https://doi.org/10.1016/j.ymeth.2021.03.018}{\doiprefix
  \detokenize{https://doi.org/10.1016/j.ymeth.2021.03.018}}.

\bibitem[{Apic et~al.(2001)Apic, Gordana and Gough, Julian and Teichmann, Sarah
  A}]{Apic2001}
\textbf{\color{eLifeMediumGrey} Apic G}, Gough J, Teichmann SA.
\newblock {Domain combinations in archaeal, eubacterial and eukaryotic
  proteomes11Edited by G. von Heijne}.
\newblock Journal of Molecular Biology.  2001; 310(2):311--325.
\newblock
  \urlprefix\url{https://www.sciencedirect.com/science/article/pii/S002228360194776X},
  \href{https://doi.org/10.1006/jmbi.2001.4776}{\doiprefix
  \detokenize{https://doi.org/10.1006/jmbi.2001.4776}}.

\bibitem[{Banani et~al.(2017)Banani, Salman F and Lee, Hyun O and Hyman,
  Anthony A and Rosen, Michael K}]{Banani2017}
\textbf{\color{eLifeMediumGrey} Banani SF}, Lee HO, Hyman AA, Rosen MK.
\newblock {Biomolecular condensates: organizers of cellular biochemistry}.
\newblock Nature Reviews Molecular Cell Biology.  2017; 18(5):285--298.
\newblock \urlprefix\url{https://doi.org/10.1038/nrm.2017.7},
  \href{10.1038/nrm.2017.7}{\doiprefix \detokenize{10.1038/nrm.2017.7}}.

\bibitem[{Beauchamp et~al.(2012)Beauchamp, Kyle A and Lin, Yu-Shan and Das,
  Rhiju and Pande, Vijay S}]{Beauchamp2012}
\textbf{\color{eLifeMediumGrey} Beauchamp KA}, Lin YS, Das R, Pande VS.
\newblock {Are Protein Force Fields Getting Better? A Systematic Benchmark on
  524 Diverse NMR Measurements}.
\newblock Journal of Chemical Theory and Computation.  2012 apr;
  8(4):1409--1414.
\newblock \urlprefix\url{https://doi.org/10.1021/ct2007814},
  \href{10.1021/ct2007814}{\doiprefix \detokenize{10.1021/ct2007814}}.

\bibitem[{Benayad et~al.(2021)Benayad, Zakarya and {Von B{\"{u}}low},
  S{\"{o}}ren and Stelzl, Lukas S. and Hummer, Gerhard}]{Benayad2021}
\textbf{\color{eLifeMediumGrey} Benayad Z}, {Von B{\"{u}}low} S, Stelzl LS,
  Hummer G.
\newblock {Simulation of FUS Protein Condensates with an Adapted Coarse-Grained
  Model}.
\newblock Journal of Chemical Theory and Computation.  2021; 17(1):525--537.
\newblock \href{10.1021/acs.jctc.0c01064}{\doiprefix
  \detokenize{10.1021/acs.jctc.0c01064}}.

\bibitem[{Berg et~al.(2018)Berg, Andrej and Kukharenko, Oleksandra and
  Scheffner, Martin and Peter, Christine}]{Berg2018}
\textbf{\color{eLifeMediumGrey} Berg A}, Kukharenko O, Scheffner M, Peter C.
\newblock {Towards a molecular basis of ubiquitin signaling: A dual-scale
  simulation study of ubiquitin dimers}.
\newblock PLoS Computational Biology.  2018; 14(11):1--14.
\newblock \href{10.1371/journal.pcbi.1006589}{\doiprefix
  \detokenize{10.1371/journal.pcbi.1006589}}.

\bibitem[{Berg and Peter(2019)Berg, Andrej and Peter,
  Christine}]{doi:10.1098/rsfs.2018.0062}
\textbf{\color{eLifeMediumGrey} Berg A}, Peter C.
\newblock {Simulating and analysing configurational landscapes of
  protein-protein contact formation}.
\newblock Interface Focus.  2019; 9(3):20180062.
\newblock
  \urlprefix\url{https://royalsocietypublishing.org/doi/abs/10.1098/rsfs.2018.0062},
  \href{10.1098/rsfs.2018.0062}{\doiprefix
  \detokenize{10.1098/rsfs.2018.0062}}.

\bibitem[{Bernad{\'{o}} et~al.(2007)Bernad{\'{o}}, Pau and Mylonas, Efstratios
  and Petoukhov, Maxim V. and Blackledge, Martin and Svergun, Dmitri
  I.}]{Bernado2007}
\textbf{\color{eLifeMediumGrey} Bernad{\'{o}} P}, Mylonas E, Petoukhov MV,
  Blackledge M, Svergun DI.
\newblock {Structural characterization of flexible proteins using small-angle
  X-ray scattering}.
\newblock Journal of the American Chemical Society.  2007; 129(17):5656--5664.
\newblock \href{10.1021/ja069124n}{\doiprefix \detokenize{10.1021/ja069124n}}.

\bibitem[{Bernad{\'{o}} and Svergun(2012)Bernad{\'{o}}, Pau and Svergun, Dmitri
  I.}]{Bernado2012}
\textbf{\color{eLifeMediumGrey} Bernad{\'{o}} P}, Svergun DI.
\newblock {Structural analysis of intrinsically disordered proteins by
  small-angle X-ray scattering}.
\newblock Molecular BioSystems.  2012; 8(1):151--167.
\newblock \href{10.1039/c1mb05275f}{\doiprefix
  \detokenize{10.1039/c1mb05275f}}.

\bibitem[{Best et~al.(2014)Best, Robert B. and Zheng, Wenwei and Mittal,
  Jeetain}]{Best2014}
\textbf{\color{eLifeMediumGrey} Best RB}, Zheng W, Mittal J.
\newblock {Balanced protein-water interactions improve properties of disordered
  proteins and non-specific protein association}.
\newblock Journal of Chemical Theory and Computation.  2014; 10(11):5113--5124.
\newblock \href{10.1021/ct500569b}{\doiprefix \detokenize{10.1021/ct500569b}}.

\bibitem[{Bock and Grubm{\"{u}}ller(2021)Bock, Lars V and Grubm{\"{u}}ller,
  Helmut}]{Bock2021}
\textbf{\color{eLifeMediumGrey} Bock LV}, Grubm{\"{u}}ller H.
\newblock {Effects of cryo-EM cooling on structural ensembles}.
\newblock bioRxiv.  2021 jan; p. 2021.10.08.463658.
\newblock
  \urlprefix\url{http://biorxiv.org/content/early/2021/10/08/2021.10.08.463658.abstract},
  \href{10.1101/2021.10.08.463658}{\doiprefix
  \detokenize{10.1101/2021.10.08.463658}}.

\bibitem[{Boeynaems et~al.(2018)Boeynaems, Steven and Alberti, Simon and Fawzi,
  Nicolas L. and Mittag, Tanja and Polymenidou, Magdalini and Rousseau,
  Frederic and Schymkowitz, Joost and Shorter, James and Wolozin, Benjamin and
  {Van Den Bosch}, Ludo and Tompa, Peter and Fuxreiter, Monika}]{Boeynaems2018}
\textbf{\color{eLifeMediumGrey} Boeynaems S}, Alberti S, Fawzi NL, Mittag T,
  Polymenidou M, Rousseau F, Schymkowitz J, Shorter J, Wolozin B, {Van Den
  Bosch} L, Tompa P, Fuxreiter M.
\newblock {Protein Phase Separation: A New Phase in Cell Biology}.
\newblock Trends in Cell Biology.  2018; 28(6):420--435.
\newblock \urlprefix\url{http://dx.doi.org/10.1016/j.tcb.2018.02.004},
  \href{10.1016/j.tcb.2018.02.004}{\doiprefix
  \detokenize{10.1016/j.tcb.2018.02.004}}.

\bibitem[{Bondos et~al.(2021)Bondos, Sarah E and Dunker, A Keith and Uversky,
  Vladimir N}]{Bondos2021}
\textbf{\color{eLifeMediumGrey} Bondos SE}, Dunker AK, Uversky VN.
\newblock {On the roles of intrinsically disordered proteins and regions in
  cell communication and signaling}.
\newblock Cell Communication and Signaling.  2021; 19(1):88.
\newblock \urlprefix\url{https://doi.org/10.1186/s12964-021-00774-3},
  \href{10.1186/s12964-021-00774-3}{\doiprefix
  \detokenize{10.1186/s12964-021-00774-3}}.

\bibitem[{Bonomi et~al.(2016)Bonomi, Massimiliano and Camilloni, Carlo and
  Cavalli, Andrea and Vendruscolo, Michele}]{Bonomi2016}
\textbf{\color{eLifeMediumGrey} Bonomi M}, Camilloni C, Cavalli A, Vendruscolo
  M.
\newblock {Metainference: A Bayesian inference method for heterogeneous
  systems}.
\newblock Science advances.  2016 jan; 2(1):e1501177--e1501177.
\newblock \urlprefix\url{https://pubmed.ncbi.nlm.nih.gov/26844300
  https://www.ncbi.nlm.nih.gov/pmc/articles/PMC4737209/},
  \href{10.1126/sciadv.1501177}{\doiprefix
  \detokenize{10.1126/sciadv.1501177}}.

\bibitem[{Bonomi et~al.(2017)Bonomi, Massimiliano and Heller, Gabriella T and
  Camilloni, Carlo and Vendruscolo, Michele}]{Bonomi2017}
\textbf{\color{eLifeMediumGrey} Bonomi M}, Heller GT, Camilloni C, Vendruscolo
  M.
\newblock {Principles of protein structural ensemble determination}.
\newblock Current Opinion in Structural Biology.  2017; 42:106--116.
\newblock
  \urlprefix\url{https://www.sciencedirect.com/science/article/pii/S0959440X16302330},
  \href{https://doi.org/10.1016/j.sbi.2016.12.004}{\doiprefix
  \detokenize{https://doi.org/10.1016/j.sbi.2016.12.004}}.

\bibitem[{Bonomi et~al.(2018)Bonomi, Massimiliano and Pellarin, Riccardo and
  Vendruscolo, Michele}]{Bonomi2018}
\textbf{\color{eLifeMediumGrey} Bonomi M}, Pellarin R, Vendruscolo M.
\newblock {Simultaneous Determination of Protein Structure and Dynamics Using
  Cryo-Electron Microscopy}.
\newblock Biophysical Journal.  2018; 114(7):1604--1613.
\newblock
  \urlprefix\url{https://www.sciencedirect.com/science/article/pii/S0006349518302881},
  \href{https://doi.org/10.1016/j.bpj.2018.02.028}{\doiprefix
  \detokenize{https://doi.org/10.1016/j.bpj.2018.02.028}}.

\bibitem[{Bonomi and Vendruscolo(2019)Bonomi, Massimiliano and Vendruscolo,
  Michele}]{Bonomi2019}
\textbf{\color{eLifeMediumGrey} Bonomi M}, Vendruscolo M.
\newblock {Determination of protein structural ensembles using cryo-electron
  microscopy}.
\newblock Current Opinion in Structural Biology.  2019; 56:37--45.
\newblock
  \urlprefix\url{https://www.sciencedirect.com/science/article/pii/S0959440X18301180},
  \href{https://doi.org/10.1016/j.sbi.2018.10.006}{\doiprefix
  \detokenize{https://doi.org/10.1016/j.sbi.2018.10.006}}.

\bibitem[{Boomsma et~al.(2014{\natexlab{a}})Boomsma, Wouter and
  Ferkinghoff-Borg, Jesper and Lindorff-Larsen, Kresten}]{Boomsma2014}
\textbf{\color{eLifeMediumGrey} Boomsma W}, Ferkinghoff-Borg J, Lindorff-Larsen
  K.
\newblock {Combining Experiments and Simulations Using the Maximum Entropy
  Principle}.
\newblock PLOS Computational Biology.  2014 feb; 10(2):e1003406.
\newblock \urlprefix\url{https://doi.org/10.1371/journal.pcbi.1003406}.

\bibitem[{Boomsma et~al.(2014{\natexlab{b}})Boomsma, Wouter and Tian, Pengfei
  and Frellsen, Jes and Ferkinghoff-Borg, Jesper and Hamelryck, Thomas and
  Lindorff-Larsen, Kresten and Vendruscolo, Michele}]{boomsma2014equilibrium}
\textbf{\color{eLifeMediumGrey} Boomsma W}, Tian P, Frellsen J,
  Ferkinghoff-Borg J, Hamelryck T, Lindorff-Larsen K, Vendruscolo M.
\newblock Equilibrium simulations of proteins using molecular fragment
  replacement and NMR chemical shifts.
\newblock Proceedings of the National Academy of Sciences.  2014;
  111(38):13852--13857.

\bibitem[{Borcherds et~al.(2021)Borcherds, Wade and Bremer, Anne and Borgia,
  Madeleine B and Mittag, Tanja}]{Borcherds2021}
\textbf{\color{eLifeMediumGrey} Borcherds W}, Bremer A, Borgia MB, Mittag T.
\newblock {How do intrinsically disordered protein regions encode a driving
  force for liquid–liquid phase separation?}
\newblock Current Opinion in Structural Biology.  2021; 67:41--50.
\newblock
  \urlprefix\url{https://www.sciencedirect.com/science/article/pii/S0959440X20301500},
  \href{https://doi.org/10.1016/j.sbi.2020.09.004}{\doiprefix
  \detokenize{https://doi.org/10.1016/j.sbi.2020.09.004}}.

\bibitem[{Borgia et~al.(2018)Borgia, Alessandro and Borgia, Madeleine B and
  Bugge, Katrine and Kissling, Vera M and Heidarsson, P{\'{e}}tur O and
  Fernandes, Catarina B and Sottini, Andrea and Soranno, Andrea and Buholzer,
  Karin J and Nettels, Daniel and Kragelund, Birthe B and Best, Robert B and
  Schuler, Benjamin}]{Borgia2018}
\textbf{\color{eLifeMediumGrey} Borgia A}, Borgia MB, Bugge K, Kissling VM,
  Heidarsson PO, Fernandes CB, Sottini A, Soranno A, Buholzer KJ, Nettels D,
  Kragelund BB, Best RB, Schuler B.
\newblock {Extreme disorder in an ultrahigh-affinity protein complex}.
\newblock Nature.  2018; 555(7694):61--66.
\newblock \urlprefix\url{https://doi.org/10.1038/nature25762},
  \href{10.1038/nature25762}{\doiprefix \detokenize{10.1038/nature25762}}.

\bibitem[{Bottaro et~al.(2020)Bottaro, Sandro and Bengtsen, Tone and
  Lindorff-Larsen, Kresten}]{Bottaro2020}
\textbf{\color{eLifeMediumGrey} Bottaro S}, Bengtsen T, Lindorff-Larsen K.
\newblock {Integrating Molecular Simulation and Experimental Data: A
  Bayesian/Maximum Entropy Reweighting Approach}.
\newblock Methods in Molecular Biology.  2020; 2112:219--240.
\newblock \href{10.1007/978-1-0716-0270-6_15}{\doiprefix
  \detokenize{10.1007/978-1-0716-0270-6_15}}.

\bibitem[{Bottaro and Lindorff-Larsen(2018)Bottaro, Sandro and Lindorff-Larsen,
  Kresten}]{Bottaro2018}
\textbf{\color{eLifeMediumGrey} Bottaro S}, Lindorff-Larsen K.
\newblock {Biophysical experiments and biomolecular simulations: A perfect
  match?}
\newblock Science.  2018 jul; 361(6400):355 LP -- 360.
\newblock
  \urlprefix\url{http://science.sciencemag.org/content/361/6400/355.abstract},
  \href{10.1126/science.aat4010}{\doiprefix
  \detokenize{10.1126/science.aat4010}}.

\bibitem[{Braun et~al.(2019)Braun, Efrem and Gilmer, Justin and Mayes, Heather
  B. and Mobley, David L. and Monroe, Jacob I. and Prasad, Samarjeet and
  Zuckerman, Daniel M.}]{Braun2019}
\textbf{\color{eLifeMediumGrey} Braun E}, Gilmer J, Mayes HB, Mobley DL, Monroe
  JI, Prasad S, Zuckerman DM.
\newblock {Best Practices for Foundations in Molecular Simulations [Article
  v1.0]}.
\newblock Living Journal of Computational Molecular Science.  2019; 1(1):1--28.
\newblock \href{10.33011/livecoms.1.1.5957}{\doiprefix
  \detokenize{10.33011/livecoms.1.1.5957}}.

\bibitem[{Br{\"u}schweiler et~al.(1992)Br{\"u}schweiler, R and Roux, B and
  Blackledge, M and Griesinger, C and Karplus, M and Ernst, R
  R}]{Brueschweiler1992}
\textbf{\color{eLifeMediumGrey} Br{\"u}schweiler R}, Roux B, Blackledge M,
  Griesinger C, Karplus M, Ernst RR.
\newblock {Influence of rapid intramolecular motion on NMR cross-relaxation
  rates. A molecular dynamics study of antamanide in solution}.
\newblock Journal of the American Chemical Society.  1992 mar;
  114(7):2289--2302.
\newblock \urlprefix\url{https://doi.org/10.1021/ja00033a002},
  \href{10.1021/ja00033a002}{\doiprefix \detokenize{10.1021/ja00033a002}}.

\bibitem[{Cavalli et~al.(2013)Cavalli, Andrea and Camilloni, Carlo and
  Vendruscolo, Michele}]{Cavalli2013}
\textbf{\color{eLifeMediumGrey} Cavalli A}, Camilloni C, Vendruscolo M.
\newblock {Molecular dynamics simulations with replica-averaged structural
  restraints generate structural ensembles according to the maximum entropy
  principle}.
\newblock The Journal of Chemical Physics.  2013 mar; 138(9):94112.
\newblock \urlprefix\url{https://doi.org/10.1063/1.4793625},
  \href{10.1063/1.4793625}{\doiprefix \detokenize{10.1063/1.4793625}}.

\bibitem[{Chen et~al.(2019)Chen, Chin-Yu and Chang, Yuan-Chih and Lin, Bo-Lin
  and Huang, Chun-Hsiang and Tsai, Ming-Daw}]{Chen2019}
\textbf{\color{eLifeMediumGrey} Chen CY}, Chang YC, Lin BL, Huang CH, Tsai MD.
\newblock {Temperature-Resolved Cryo-EM Uncovers Structural Bases of
  Temperature-Dependent Enzyme Functions}.
\newblock Journal of the American Chemical Society.  2019 dec;
  141(51):19983--19987.
\newblock \urlprefix\url{https://doi.org/10.1021/jacs.9b10687},
  \href{10.1021/jacs.9b10687}{\doiprefix \detokenize{10.1021/jacs.9b10687}}.

\bibitem[{Cho et~al.(2021)Cho, Hyun Sun and Schotte, Friedrich and Stadnytskyi,
  Valentyn and Anfinrud, Philip}]{Cho2021}
\textbf{\color{eLifeMediumGrey} Cho HS}, Schotte F, Stadnytskyi V, Anfinrud P.
\newblock {Time-resolved X-ray scattering studies of proteins}.
\newblock Current Opinion in Structural Biology.  2021; 70:99--107.
\newblock
  \urlprefix\url{https://www.sciencedirect.com/science/article/pii/S0959440X21000646},
  \href{https://doi.org/10.1016/j.sbi.2021.05.002}{\doiprefix
  \detokenize{https://doi.org/10.1016/j.sbi.2021.05.002}}.

\bibitem[{Choi et~al.(2020)Choi, Jeong-Mo and Holehouse, Alex S and Pappu,
  Rohit V}]{Choi2020}
\textbf{\color{eLifeMediumGrey} Choi JM}, Holehouse AS, Pappu RV.
\newblock {Physical Principles Underlying the Complex Biology of Intracellular
  Phase Transitions}.
\newblock Annual Review of Biophysics.  2020 may; 49(1):107--133.
\newblock
  \urlprefix\url{https://doi.org/10.1146/annurev-biophys-121219-081629},
  \href{10.1146/annurev-biophys-121219-081629}{\doiprefix
  \detokenize{10.1146/annurev-biophys-121219-081629}}.

\bibitem[{Choi and Pappu(2019)Choi, Jeong-Mo and Pappu, Rohit V}]{Choi2019}
\textbf{\color{eLifeMediumGrey} Choi JM}, Pappu RV.
\newblock {Improvements to the ABSINTH Force Field for Proteins Based on
  Experimentally Derived Amino Acid Specific Backbone Conformational
  Statistics}.
\newblock Journal of Chemical Theory and Computation.  2019 feb;
  15(2):1367--1382.
\newblock \urlprefix\url{https://doi.org/10.1021/acs.jctc.8b00573},
  \href{10.1021/acs.jctc.8b00573}{\doiprefix
  \detokenize{10.1021/acs.jctc.8b00573}}.

\bibitem[{Clore and Iwahara(2009)Clore, G Marius and Iwahara,
  Junji}]{Clore2009}
\textbf{\color{eLifeMediumGrey} Clore GM}, Iwahara J.
\newblock {Theory, Practice, and Applications of Paramagnetic Relaxation
  Enhancement for the Characterization of Transient Low-Population States of
  Biological Macromolecules and Their Complexes}.
\newblock Chemical Reviews.  2009 sep; 109(9):4108--4139.
\newblock \urlprefix\url{https://doi.org/10.1021/cr900033p},
  \href{10.1021/cr900033p}{\doiprefix \detokenize{10.1021/cr900033p}}.

\bibitem[{Cossio and Hummer(2013)Cossio, Pilar and Hummer,
  Gerhard}]{Cossio2013}
\textbf{\color{eLifeMediumGrey} Cossio P}, Hummer G.
\newblock {Bayesian analysis of individual electron microscopy images: Towards
  structures of dynamic and heterogeneous biomolecular assemblies}.
\newblock Journal of Structural Biology.  2013; 184(3):427--437.
\newblock
  \urlprefix\url{https://www.sciencedirect.com/science/article/pii/S1047847713002712},
  \href{https://doi.org/10.1016/j.jsb.2013.10.006}{\doiprefix
  \detokenize{https://doi.org/10.1016/j.jsb.2013.10.006}}.

\bibitem[{Cragnell et~al.(2016)Cragnell, Carolina and Durand, Dominique and
  Cabane, Bernard and Skep{\"{o}}, Marie}]{Cragnell2016}
\textbf{\color{eLifeMediumGrey} Cragnell C}, Durand D, Cabane B, Skep{\"{o}} M.
\newblock {Coarse-grained modeling of the intrinsically disordered protein
  Histatin 5 in solution: Monte Carlo simulations in combination with SAXS}.
\newblock Proteins: Structure, Function, and Bioinformatics.  2016 jun;
  84(6):777--791.
\newblock \urlprefix\url{https://doi.org/10.1002/prot.25025},
  \href{https://doi.org/10.1002/prot.25025}{\doiprefix
  \detokenize{https://doi.org/10.1002/prot.25025}}.

\bibitem[{Cragnell et~al.(2018)Cragnell, Carolina and Rieloff, Ellen and
  Skep{\"{o}}, Marie}]{Cragnell2018}
\textbf{\color{eLifeMediumGrey} Cragnell C}, Rieloff E, Skep{\"{o}} M.
\newblock {Utilizing Coarse-Grained Modeling and Monte Carlo Simulations to
  Evaluate the Conformational Ensemble of Intrinsically Disordered Proteins and
  Regions}.
\newblock Journal of Molecular Biology.  2018; 430(16):2478--2492.
\newblock
  \urlprefix\url{https://www.sciencedirect.com/science/article/pii/S0022283618301347},
  \href{https://doi.org/10.1016/j.jmb.2018.03.006}{\doiprefix
  \detokenize{https://doi.org/10.1016/j.jmb.2018.03.006}}.

\bibitem[{Das et~al.(2018{\natexlab{a}})Das, Suman and Amin, Alan N and Lin,
  Yi-Hsuan and Chan, Hue Sun}]{Das2018a}
\textbf{\color{eLifeMediumGrey} Das S}, Amin AN, Lin YH, Chan HS.
\newblock {Coarse-grained residue-based models of disordered protein
  condensates: utility and limitations of simple charge pattern parameters}.
\newblock Physical Chemistry Chemical Physics.  2018; 20(45):28558--28574.
\newblock \urlprefix\url{http://dx.doi.org/10.1039/C8CP05095C},
  \href{10.1039/C8CP05095C}{\doiprefix \detokenize{10.1039/C8CP05095C}}.

\bibitem[{Das et~al.(2018{\natexlab{b}})Das, Suman and Eisen, Adam and Lin,
  Yi-Hsuan and Chan, Hue Sun}]{Das2018}
\textbf{\color{eLifeMediumGrey} Das S}, Eisen A, Lin YH, Chan HS.
\newblock {A Lattice Model of Charge-Pattern-Dependent Polyampholyte Phase
  Separation}.
\newblock The Journal of Physical Chemistry B.  2018 may; 122(21):5418--5431.
\newblock \urlprefix\url{https://doi.org/10.1021/acs.jpcb.7b11723},
  \href{10.1021/acs.jpcb.7b11723}{\doiprefix
  \detokenize{10.1021/acs.jpcb.7b11723}}.

\bibitem[{Delaforge et~al.(2016)Delaforge, Elise and Milles, Sigrid and Huang,
  Jie-rong and Bouvier, Denis and Jensen, Malene Ringkj{\o}bing and Sattler,
  Michael and Hart, Darren J and Blackledge, Martin}]{10.3389/fmolb.2016.00054}
\textbf{\color{eLifeMediumGrey} Delaforge E}, Milles S, Huang Jr, Bouvier D,
  Jensen MR, Sattler M, Hart DJ, Blackledge M.
\newblock {Investigating the Role of Large-Scale Domain Dynamics in
  Protein-Protein Interactions}.
\newblock Frontiers in Molecular Biosciences.  2016; 3:54.
\newblock
  \urlprefix\url{https://www.frontiersin.org/article/10.3389/fmolb.2016.00054},
  \href{10.3389/fmolb.2016.00054}{\doiprefix
  \detokenize{10.3389/fmolb.2016.00054}}.

\bibitem[{Dignon et~al.(2020)Dignon, Gregory L and Best, Robert B and Mittal,
  Jeetain}]{Dignon2020}
\textbf{\color{eLifeMediumGrey} Dignon GL}, Best RB, Mittal J.
\newblock {Biomolecular Phase Separation: From Molecular Driving Forces to
  Macroscopic Properties}.
\newblock Annual Review of Physical Chemistry.  2020 apr; 71(1):53--75.
\newblock
  \urlprefix\url{https://doi.org/10.1146/annurev-physchem-071819-113553},
  \href{10.1146/annurev-physchem-071819-113553}{\doiprefix
  \detokenize{10.1146/annurev-physchem-071819-113553}}.

\bibitem[{Dignon et~al.(2018)Dignon, Gregory L and Zheng, Wenwei and Kim, Young
  C and Best, Robert B and Mittal, Jeetain}]{Dignon2018}
\textbf{\color{eLifeMediumGrey} Dignon GL}, Zheng W, Kim YC, Best RB, Mittal J.
\newblock {Sequence determinants of protein phase behavior from a
  coarse-grained model}.
\newblock PLOS Computational Biology.  2018 jan; 14(1):e1005941.
\newblock \urlprefix\url{https://doi.org/10.1371/journal.pcbi.1005941}.

\bibitem[{Dunker and Uversky(2010)Dunker, A Keith and Uversky, Vladimir
  N}]{Dunker2010}
\textbf{\color{eLifeMediumGrey} Dunker AK}, Uversky VN.
\newblock {Drugs for ‘protein clouds': targeting intrinsically disordered
  transcription factors}.
\newblock Current Opinion in Pharmacology.  2010; 10(6):782--788.
\newblock
  \urlprefix\url{https://www.sciencedirect.com/science/article/pii/S1471489210001438},
  \href{https://doi.org/10.1016/j.coph.2010.09.005}{\doiprefix
  \detokenize{https://doi.org/10.1016/j.coph.2010.09.005}}.

\bibitem[{Ekman et~al.(2005)Ekman, Diana and Bj{\"{o}}rklund, {\AA}sa K and
  Frey-Sk{\"{o}}tt, Johannes and Elofsson, Arne}]{Ekman2005}
\textbf{\color{eLifeMediumGrey} Ekman D}, Bj{\"{o}}rklund {\AA}K,
  Frey-Sk{\"{o}}tt J, Elofsson A.
\newblock {Multi-domain Proteins in the Three Kingdoms of Life: Orphan Domains
  and Other Unassigned Regions}.
\newblock Journal of Molecular Biology.  2005; 348(1):231--243.
\newblock
  \urlprefix\url{https://www.sciencedirect.com/science/article/pii/S0022283605001609},
  \href{https://doi.org/10.1016/j.jmb.2005.02.007}{\doiprefix
  \detokenize{https://doi.org/10.1016/j.jmb.2005.02.007}}.

\bibitem[{Fischer et~al.(2010)Fischer, Niels and Konevega, Andrey L and
  Wintermeyer, Wolfgang and Rodnina, Marina V and Stark, Holger}]{Fischer2010}
\textbf{\color{eLifeMediumGrey} Fischer N}, Konevega AL, Wintermeyer W, Rodnina
  MV, Stark H.
\newblock {Ribosome dynamics and tRNA movement by time-resolved electron
  cryomicroscopy}.
\newblock Nature.  2010; 466(7304):329--333.
\newblock \urlprefix\url{https://doi.org/10.1038/nature09206},
  \href{10.1038/nature09206}{\doiprefix \detokenize{10.1038/nature09206}}.

\bibitem[{Fuertes et~al.(2017)Fuertes, Gustavo and Banterle, Niccol{\`{o}} and
  Ruff, Kiersten M and Chowdhury, Aritra and Mercadante, Davide and Koehler,
  Christine and Kachala, Michael and {Estrada Girona}, Gemma and Milles, Sigrid
  and Mishra, Ankur and Onck, Patrick R and Gr{\"{a}}ter, Frauke and
  Esteban-Mart{\'{i}}n, Santiago and Pappu, Rohit V and Svergun, Dmitri I and
  Lemke, Edward A}]{Fuertes2017}
\textbf{\color{eLifeMediumGrey} Fuertes G}, Banterle N, Ruff KM, Chowdhury A,
  Mercadante D, Koehler C, Kachala M, {Estrada Girona} G, Milles S, Mishra A,
  Onck PR, Gr{\"{a}}ter F, Esteban-Mart{\'{i}}n S, Pappu RV, Svergun DI, Lemke
  EA.
\newblock {Decoupling of size and shape fluctuations in heteropolymeric
  sequences reconciles discrepancies in SAXS vs. FRET measurements}.
\newblock Proceedings of the National Academy of Sciences.  2017 aug;
  114(31):E6342 LP -- E6351.
\newblock \urlprefix\url{http://www.pnas.org/content/114/31/E6342.abstract},
  \href{10.1073/pnas.1704692114}{\doiprefix
  \detokenize{10.1073/pnas.1704692114}}.

\bibitem[{Gkeka et~al.(2020)Gkeka, Paraskevi and Stoltz, Gabriel and {Barati
  Farimani}, Amir and Belkacemi, Zineb and Ceriotti, Michele and Chodera, John
  D and Dinner, Aaron R and Ferguson, Andrew L and Maillet, Jean-Bernard and
  Minoux, Herv{\'{e}} and Peter, Christine and Pietrucci, Fabio and Silveira,
  Ana and Tkatchenko, Alexandre and Trstanova, Zofia and Wiewiora, Rafal and
  Leli{\`{e}}vre, Tony}]{Gkeka2020}
\textbf{\color{eLifeMediumGrey} Gkeka P}, Stoltz G, {Barati Farimani} A,
  Belkacemi Z, Ceriotti M, Chodera JD, Dinner AR, Ferguson AL, Maillet JB,
  Minoux H, Peter C, Pietrucci F, Silveira A, Tkatchenko A, Trstanova Z,
  Wiewiora R, Leli{\`{e}}vre T.
\newblock {Machine Learning Force Fields and Coarse-Grained Variables in
  Molecular Dynamics: Application to Materials and Biological Systems}.
\newblock Journal of Chemical Theory and Computation.  2020 aug;
  16(8):4757--4775.
\newblock \urlprefix\url{https://doi.org/10.1021/acs.jctc.0c00355},
  \href{10.1021/acs.jctc.0c00355}{\doiprefix
  \detokenize{10.1021/acs.jctc.0c00355}}.

\bibitem[{Gomes et~al.(2020)Gomes, Gregory Neal W. and Krzeminski,
  Micka{\"{e}}l and Namini, Ashley and Martin, Erik W. and Mittag, Tanja and
  Head-Gordon, Teresa and Forman-Kay, Julie D. and Gradinaru, Claudiu
  C.}]{Gomes2020}
\textbf{\color{eLifeMediumGrey} Gomes GNW}, Krzeminski M, Namini A, Martin EW,
  Mittag T, Head-Gordon T, Forman-Kay JD, Gradinaru CC.
\newblock {Conformational Ensembles of an Intrinsically Disordered Protein
  Consistent with NMR, SAXS, and Single-Molecule FRET}.
\newblock Journal of the American Chemical Society.  2020;
  142(37):15697--15710.
\newblock \href{10.1021/jacs.0c02088}{\doiprefix
  \detokenize{10.1021/jacs.0c02088}}.

\bibitem[{Grudinin et~al.(2017)Grudinin, Sergei and Garkavenko, Maria and
  Kazennov, Andrei}]{Grudinin2017}
\textbf{\color{eLifeMediumGrey} Grudinin S}, Garkavenko M, Kazennov A.
\newblock {Pepsi-SAXS: An adaptive method for rapid and accurate computation of
  small-angle X-ray scattering profiles}.
\newblock Acta Crystallographica Section D: Structural Biology.  2017;
  73(5):449--464.
\newblock \href{10.1107/S2059798317005745}{\doiprefix
  \detokenize{10.1107/S2059798317005745}}.

\bibitem[{Gumerov et~al.(2012)Gumerov, Nail A and Berlin, Konstantin and
  Fushman, David and Duraiswami, Ramani}]{Gumerov2012}
\textbf{\color{eLifeMediumGrey} Gumerov NA}, Berlin K, Fushman D, Duraiswami R.
\newblock {A hierarchical algorithm for fast Debye summation with applications
  to small angle scattering.}
\newblock Journal of computational chemistry.  2012 sep; 33(25):1981--1996.
\newblock \href{10.1002/jcc.23025}{\doiprefix \detokenize{10.1002/jcc.23025}}.

\bibitem[{Han et~al.(2011)Han, Beomsoo and Liu, Yifeng and Ginzinger, Simon W
  and Wishart, David S}]{Han2011}
\textbf{\color{eLifeMediumGrey} Han B}, Liu Y, Ginzinger SW, Wishart DS.
\newblock {SHIFTX2: significantly improved protein chemical shift prediction}.
\newblock Journal of Biomolecular NMR.  2011; 50(1):43.
\newblock \urlprefix\url{https://doi.org/10.1007/s10858-011-9478-4},
  \href{10.1007/s10858-011-9478-4}{\doiprefix
  \detokenize{10.1007/s10858-011-9478-4}}.

\bibitem[{Haris(2013)Haris, Parvez I}]{Haris2013}
\textbf{\color{eLifeMediumGrey} Haris PI}.
\newblock {Infrared Spectroscopy of Protein Structure BT}.
\newblock In: Roberts GCK, editor. \emph{{Encyclopedia of Biophysics}} Berlin,
  Heidelberg: Springer Berlin Heidelberg; 2013.p. 1095--1106.
\newblock \urlprefix\url{https://doi.org/10.1007/978-3-642-16712-6_135},
  \href{10.1007/978-3-642-16712-6_135}{\doiprefix
  \detokenize{10.1007/978-3-642-16712-6_135}}.

\bibitem[{Heller et~al.(2017)Heller, Gabriella T and Aprile, Francesco A and
  Bonomi, Massimiliano and Camilloni, Carlo and {De Simone}, Alfonso and
  Vendruscolo, Michele}]{Heller2017}
\textbf{\color{eLifeMediumGrey} Heller GT}, Aprile FA, Bonomi M, Camilloni C,
  {De Simone} A, Vendruscolo M.
\newblock {Sequence Specificity in the Entropy-Driven Binding of a Small
  Molecule and a Disordered Peptide}.
\newblock Journal of Molecular Biology.  2017; 429(18):2772--2779.
\newblock
  \urlprefix\url{https://www.sciencedirect.com/science/article/pii/S0022283617303534},
  \href{https://doi.org/10.1016/j.jmb.2017.07.016}{\doiprefix
  \detokenize{https://doi.org/10.1016/j.jmb.2017.07.016}}.

\bibitem[{Heller et~al.(2020)Heller, Gabriella T and Aprile, Francesco A and
  Michaels, Thomas C T and Limbocker, Ryan and Perni, Michele and Ruggeri,
  Francesco Simone and Mannini, Benedetta and L{\"{o}}hr, Thomas and Bonomi,
  Massimiliano and Camilloni, Carlo and Simone, Alfonso De and Felli, Isabella
  C and Pierattelli, Roberta and Knowles, Tuomas P J and Dobson, Christopher M
  and Vendruscolo, Michele}]{doi:10.1126/sciadv.abb5924}
\textbf{\color{eLifeMediumGrey} Heller GT}, Aprile FA, Michaels TCT, Limbocker
  R, Perni M, Ruggeri FS, Mannini B, L{\"{o}}hr T, Bonomi M, Camilloni C,
  Simone AD, Felli IC, Pierattelli R, Knowles TPJ, Dobson CM, Vendruscolo M.
\newblock {Small-molecule sequestration of amyloid-beta as a drug discovery
  strategy for Alzheimer's disease}.
\newblock Science Advances.  2020; 6(45):eabb5924.
\newblock \href{10.1126/sciadv.abb5924}{\doiprefix
  \detokenize{10.1126/sciadv.abb5924}}.

\bibitem[{Heller et~al.(2018)Heller, Gabriella T and Bonomi, Massimiliano and
  Vendruscolo, Michele}]{Heller2018}
\textbf{\color{eLifeMediumGrey} Heller GT}, Bonomi M, Vendruscolo M.
\newblock {Structural Ensemble Modulation upon Small-Molecule Binding to
  Disordered Proteins}.
\newblock Journal of Molecular Biology.  2018; 430(16):2288--2292.
\newblock
  \urlprefix\url{https://www.sciencedirect.com/science/article/pii/S0022283618301621},
  \href{https://doi.org/10.1016/j.jmb.2018.03.015}{\doiprefix
  \detokenize{https://doi.org/10.1016/j.jmb.2018.03.015}}.

\bibitem[{Heller et~al.(2015)Heller, Gabriella T and Sormanni, Pietro and
  Vendruscolo, Michele}]{Heller2015}
\textbf{\color{eLifeMediumGrey} Heller GT}, Sormanni P, Vendruscolo M.
\newblock {Targeting disordered proteins with small molecules using entropy}.
\newblock Trends in Biochemical Sciences.  2015 sep; 40(9):491--496.
\newblock \urlprefix\url{https://doi.org/10.1016/j.tibs.2015.07.004},
  \href{10.1016/j.tibs.2015.07.004}{\doiprefix
  \detokenize{10.1016/j.tibs.2015.07.004}}.

\bibitem[{Henriques et~al.(2018)Henriques, Jo{\~{a}}o and Arleth, Lise and
  Lindorff-Larsen, Kresten and Skep{\"{o}}, Marie}]{Henriques2018}
\textbf{\color{eLifeMediumGrey} Henriques J}, Arleth L, Lindorff-Larsen K,
  Skep{\"{o}} M.
\newblock {On the Calculation of SAXS Profiles of Folded and Intrinsically
  Disordered Proteins from Computer Simulations}.
\newblock Journal of Molecular Biology.  2018; 430(16):2521--2539.
\newblock
  \urlprefix\url{https://www.sciencedirect.com/science/article/pii/S0022283618301232},
  \href{https://doi.org/10.1016/j.jmb.2018.03.002}{\doiprefix
  \detokenize{https://doi.org/10.1016/j.jmb.2018.03.002}}.

\bibitem[{Henriques et~al.(2015)Henriques, Jo{\~{a}}o and Cragnell, Carolina
  and Skep{\"{o}}, Marie}]{Henriques2015}
\textbf{\color{eLifeMediumGrey} Henriques J}, Cragnell C, Skep{\"{o}} M.
\newblock {Molecular Dynamics Simulations of Intrinsically Disordered Proteins:
  Force Field Evaluation and Comparison with Experiment}.
\newblock Journal of Chemical Theory and Computation.  2015 jul;
  11(7):3420--3431.
\newblock \urlprefix\url{https://doi.org/10.1021/ct501178z},
  \href{10.1021/ct501178z}{\doiprefix \detokenize{10.1021/ct501178z}}.

\bibitem[{Hollingsworth and Dror(2018)Hollingsworth, Scott A and Dror, Ron
  O}]{Hollingsworth2018}
\textbf{\color{eLifeMediumGrey} Hollingsworth SA}, Dror RO.
\newblock {Molecular Dynamics Simulation for All}.
\newblock Neuron.  2018; 99(6):1129--1143.
\newblock
  \urlprefix\url{https://www.sciencedirect.com/science/article/pii/S0896627318306846},
  \href{https://doi.org/10.1016/j.neuron.2018.08.011}{\doiprefix
  \detokenize{https://doi.org/10.1016/j.neuron.2018.08.011}}.

\bibitem[{Hu et~al.(2010)Hu, Kan-Nian and Yau, Wai-Ming and Tycko,
  Robert}]{hu2010detection}
\textbf{\color{eLifeMediumGrey} Hu KN}, Yau WM, Tycko R.
\newblock Detection of a transient intermediate in a rapid protein folding
  process by solid-state nuclear magnetic resonance.
\newblock Journal of the American Chemical Society.  2010; 132(1):24--25.

\bibitem[{Huang et~al.(2017)Huang, Jing and Rauscher, Sarah and Nawrocki,
  Grzegorz and Ran, Ting and Feig, Michael and de Groot, Bert L and
  Grubm{\"{u}}ller, Helmut and MacKerell, Alexander D}]{Huang2017}
\textbf{\color{eLifeMediumGrey} Huang J}, Rauscher S, Nawrocki G, Ran T, Feig
  M, de~Groot BL, Grubm{\"{u}}ller H, MacKerell AD.
\newblock {CHARMM36m: an improved force field for folded and intrinsically
  disordered proteins}.
\newblock Nature Methods.  2017; 14(1):71--73.
\newblock \urlprefix\url{https://doi.org/10.1038/nmeth.4067},
  \href{10.1038/nmeth.4067}{\doiprefix \detokenize{10.1038/nmeth.4067}}.

\bibitem[{Huang et~al.(2020)Huang, Lei and Wright, Michelle and Yang, Sichun
  and Blachowicz, Lydia and Makowski, Lee and Roux, Beno{\^{i}}t}]{Huang2020}
\textbf{\color{eLifeMediumGrey} Huang L}, Wright M, Yang S, Blachowicz L,
  Makowski L, Roux B.
\newblock {Glycine substitution in SH3-SH2 connector of Hck tyrosine kinase
  causes population shift from assembled to disassembled state}.
\newblock Biochimica et Biophysica Acta (BBA) - General Subjects.  2020;
  1864(7):129604.
\newblock
  \urlprefix\url{https://www.sciencedirect.com/science/article/pii/S0304416520301161},
  \href{https://doi.org/10.1016/j.bbagen.2020.129604}{\doiprefix
  \detokenize{https://doi.org/10.1016/j.bbagen.2020.129604}}.

\bibitem[{Hub(2018)Hub, Jochen S}]{Hub2018}
\textbf{\color{eLifeMediumGrey} Hub JS}.
\newblock {Interpreting solution X-ray scattering data using molecular
  simulations}.
\newblock Current Opinion in Structural Biology.  2018; 49:18--26.
\newblock
  \urlprefix\url{https://www.sciencedirect.com/science/article/pii/S0959440X17300891},
  \href{https://doi.org/10.1016/j.sbi.2017.11.002}{\doiprefix
  \detokenize{https://doi.org/10.1016/j.sbi.2017.11.002}}.

\bibitem[{Hummer and K{\"{o}}finger(2015)Hummer, Gerhard and K{\"{o}}finger,
  J{\"{u}}rgen}]{Hummer2015}
\textbf{\color{eLifeMediumGrey} Hummer G}, K{\"{o}}finger J.
\newblock {Bayesian ensemble refinement by replica simulations and
  reweighting}.
\newblock The Journal of Chemical Physics.  2015 dec; 143(24):243150.
\newblock \urlprefix\url{https://doi.org/10.1063/1.4937786},
  \href{10.1063/1.4937786}{\doiprefix \detokenize{10.1063/1.4937786}}.

\bibitem[{Ing{\'{o}}lfsson et~al.(2014)Ing{\'{o}}lfsson, Helgi I. and Lopez,
  Cesar A. and Uusitalo, Jaakko J. and de Jong, Djurre H. and Gopal, Srinivasa
  M. and Periole, Xavier and Marrink, Siewert J.}]{Ingolfsson2014}
\textbf{\color{eLifeMediumGrey} Ing{\'{o}}lfsson HI}, Lopez CA, Uusitalo JJ,
  de~Jong DH, Gopal SM, Periole X, Marrink SJ.
\newblock {The power of coarse graining in biomolecular simulations}.
\newblock Wiley Interdisciplinary Reviews: Computational Molecular Science.
  2014; 4(3):225--248.
\newblock \href{10.1002/wcms.1169}{\doiprefix \detokenize{10.1002/wcms.1169}}.

\bibitem[{Jaynes(1957)Jaynes, E T}]{PhysRev.106.620}
\textbf{\color{eLifeMediumGrey} Jaynes ET}.
\newblock {Information Theory and Statistical Mechanics}.
\newblock Phys Rev.  1957 may; 106(4):620--630.
\newblock \urlprefix\url{https://link.aps.org/doi/10.1103/PhysRev.106.620},
  \href{10.1103/PhysRev.106.620}{\doiprefix
  \detokenize{10.1103/PhysRev.106.620}}.

\bibitem[{Jephthah et~al.(2021)Jephthah, St{\'e}phanie and Pesce, Francesco and
  Lindorff-Larsen, Kresten and Skep{\"o}, Marie}]{jephthah2021force}
\textbf{\color{eLifeMediumGrey} Jephthah S}, Pesce F, Lindorff-Larsen K,
  Skep{\"o} M.
\newblock Force Field Effects in Simulations of Flexible Peptides with Varying
  Polyproline II Propensity.
\newblock Journal of chemical theory and computation.  2021; 17(10):6634--6646.

\bibitem[{Jumper et~al.(2021)Jumper, John and Evans, Richard and Pritzel,
  Alexander and Green, Tim and Figurnov, Michael and Ronneberger, Olaf and
  Tunyasuvunakool, Kathryn and Bates, Russ and {\v{Z}}{\'{i}}dek, Augustin and
  Potapenko, Anna and Bridgland, Alex and Meyer, Clemens and Kohl, Simon A A
  and Ballard, Andrew J and Cowie, Andrew and Romera-Paredes, Bernardino and
  Nikolov, Stanislav and Jain, Rishub and Adler, Jonas and Back, Trevor and
  Petersen, Stig and Reiman, David and Clancy, Ellen and Zielinski, Michal and
  Steinegger, Martin and Pacholska, Michalina and Berghammer, Tamas and
  Bodenstein, Sebastian and Silver, David and Vinyals, Oriol and Senior, Andrew
  W and Kavukcuoglu, Koray and Kohli, Pushmeet and Hassabis,
  Demis}]{Jumper2021}
\textbf{\color{eLifeMediumGrey} Jumper J}, Evans R, Pritzel A, Green T,
  Figurnov M, Ronneberger O, Tunyasuvunakool K, Bates R, {\v{Z}}{\'{i}}dek A,
  Potapenko A, Bridgland A, Meyer C, Kohl SAA, Ballard AJ, Cowie A,
  Romera-Paredes B, Nikolov S, Jain R, Adler J, Back T, et~al.
\newblock {Highly accurate protein structure prediction with AlphaFold}.
\newblock Nature.  2021; 596(7873):583--589.
\newblock \urlprefix\url{https://doi.org/10.1038/s41586-021-03819-2},
  \href{10.1038/s41586-021-03819-2}{\doiprefix
  \detokenize{10.1038/s41586-021-03819-2}}.

\bibitem[{Jussupow et~al.(2020)Jussupow, Alexander and Messias, Ana C and
  Stehle, Ralf and Geerlof, Arie and Solbak, Sara M {\O} and Paissoni, Cristina
  and Bach, Anders and Sattler, Michael and Camilloni,
  Carlo}]{doi:10.1126/sciadv.abc3786}
\textbf{\color{eLifeMediumGrey} Jussupow A}, Messias AC, Stehle R, Geerlof A,
  Solbak SM{\O}, Paissoni C, Bach A, Sattler M, Camilloni C.
\newblock {The dynamics of linear polyubiquitin}.
\newblock Science Advances.  2020; 6(42):eabc3786.
\newblock \href{10.1126/sciadv.abc3786}{\doiprefix
  \detokenize{10.1126/sciadv.abc3786}}.

\bibitem[{Kalinin et~al.(2012)Kalinin, Stanislav and Peulen, Thomas and
  Sindbert, Simon and Rothwell, Paul J and Berger, Sylvia and Restle, Tobias
  and Goody, Roger S and Gohlke, Holger and Seidel, Claus A M}]{Kalinin2012}
\textbf{\color{eLifeMediumGrey} Kalinin S}, Peulen T, Sindbert S, Rothwell PJ,
  Berger S, Restle T, Goody RS, Gohlke H, Seidel CAM.
\newblock {A toolkit and benchmark study for FRET-restrained high-precision
  structural modeling}.
\newblock Nature Methods.  2012; 9(12):1218--1225.
\newblock \urlprefix\url{https://doi.org/10.1038/nmeth.2222},
  \href{10.1038/nmeth.2222}{\doiprefix \detokenize{10.1038/nmeth.2222}}.

\bibitem[{Kassem et~al.(2021)Kassem, Noah and Araya-Secchi, Raul and Bugge,
  Katrine and Barclay, Abigail and Steinocher, Helena and Khondker, Adree and
  Wang, Yong and Lenard, Aneta J and B{\"{u}}rck, Jochen and Sahin, Cagla and
  Ulrich, Anne S and Landreh, Michael and Pedersen, Martin Cramer and
  Rheinst{\"{a}}dter, Maikel C and Pedersen, Per Amstrup and Lindorff-Larsen,
  Kresten and Arleth, Lise and Kragelund, Birthe
  B}]{doi:10.1126/sciadv.abh3805}
\textbf{\color{eLifeMediumGrey} Kassem N}, Araya-Secchi R, Bugge K, Barclay A,
  Steinocher H, Khondker A, Wang Y, Lenard AJ, B{\"{u}}rck J, Sahin C, Ulrich
  AS, Landreh M, Pedersen MC, Rheinst{\"{a}}dter MC, Pedersen PA,
  Lindorff-Larsen K, Arleth L, Kragelund BB.
\newblock {Order and disorder-An integrative structure of the full-length human
  growth hormone receptor}.
\newblock Science Advances.  2021; 7(27):eabh3805.
\newblock \href{10.1126/sciadv.abh3805}{\doiprefix
  \detokenize{10.1126/sciadv.abh3805}}.

\bibitem[{Klose et~al.(2021)Klose, Daniel and Holla, Andrea and Gmeiner,
  Christoph and Nettels, Daniel and Ritsch, Irina and Bross, Nadja and Yulikov,
  Maxim and Allain, Fr{\'{e}}d{\'{e}}ric H.-T. and Schuler, Benjamin and
  Jeschke, Gunnar}]{Klose2021}
\textbf{\color{eLifeMediumGrey} Klose D}, Holla A, Gmeiner C, Nettels D, Ritsch
  I, Bross N, Yulikov M, Allain FHT, Schuler B, Jeschke G.
\newblock {Resolving distance variations by single-molecule FRET and EPR
  spectroscopy using rotamer libraries}.
\newblock Biophysical Journal.  2021; 120(21):4842--4858.
\newblock
  \urlprefix\url{https://www.sciencedirect.com/science/article/pii/S0006349521007554},
  \href{https://doi.org/10.1016/j.bpj.2021.09.021}{\doiprefix
  \detokenize{https://doi.org/10.1016/j.bpj.2021.09.021}}.

\bibitem[{Kmiecik et~al.(2016)Kmiecik, Sebastian and Gront, Dominik and
  Kolinski, Michal and Wieteska, Lukasz and Dawid, Aleksandra Elzbieta and
  Kolinski, Andrzej}]{Kmiecik2016}
\textbf{\color{eLifeMediumGrey} Kmiecik S}, Gront D, Kolinski M, Wieteska L,
  Dawid AE, Kolinski A.
\newblock {Coarse-Grained Protein Models and Their Applications}.
\newblock Chemical Reviews.  2016 jul; 116(14):7898--7936.
\newblock \urlprefix\url{https://doi.org/10.1021/acs.chemrev.6b00163},
  \href{10.1021/acs.chemrev.6b00163}{\doiprefix
  \detokenize{10.1021/acs.chemrev.6b00163}}.

\bibitem[{K{\"{o}}finger and Hummer(2021)K{\"{o}}finger, J{\"{u}}rgen and
  Hummer, Gerhard}]{Kofinger2021}
\textbf{\color{eLifeMediumGrey} K{\"{o}}finger J}, Hummer G.
\newblock {Empirical optimization of molecular simulation force fields by
  Bayesian inference}.
\newblock European Physical Journal B.  2021; .

\bibitem[{Kohlhoff et~al.(2009)Kohlhoff, Kai J and Robustelli, Paul and
  Cavalli, Andrea and Salvatella, Xavier and Vendruscolo,
  Michele}]{Kohlhoff2009}
\textbf{\color{eLifeMediumGrey} Kohlhoff KJ}, Robustelli P, Cavalli A,
  Salvatella X, Vendruscolo M.
\newblock {Fast and Accurate Predictions of Protein NMR Chemical Shifts from
  Interatomic Distances}.
\newblock Journal of the American Chemical Society.  2009 oct;
  131(39):13894--13895.
\newblock \urlprefix\url{https://doi.org/10.1021/ja903772t},
  \href{10.1021/ja903772t}{\doiprefix \detokenize{10.1021/ja903772t}}.

\bibitem[{Kooshapur et~al.(2018)Kooshapur, Hamed and Choudhury, Nila Roy and
  Simon, Bernd and M{\"{u}}hlbauer, Max and Jussupow, Alexander and Fernandez,
  Noemi and Jones, Alisha N and Dallmann, Andre and Gabel, Frank and Camilloni,
  Carlo and Michlewski, Gracjan and Caceres, Javier F and Sattler,
  Michael}]{Kooshapur2018}
\textbf{\color{eLifeMediumGrey} Kooshapur H}, Choudhury NR, Simon B,
  M{\"{u}}hlbauer M, Jussupow A, Fernandez N, Jones AN, Dallmann A, Gabel F,
  Camilloni C, Michlewski G, Caceres JF, Sattler M.
\newblock {Structural basis for terminal loop recognition and stimulation of
  pri-miRNA-18a processing by hnRNP A1}.
\newblock Nature Communications.  2018; 9(1):2479.
\newblock \urlprefix\url{https://doi.org/10.1038/s41467-018-04871-9},
  \href{10.1038/s41467-018-04871-9}{\doiprefix
  \detokenize{10.1038/s41467-018-04871-9}}.

\bibitem[{Kragelj et~al.(2013)Kragelj, Jaka and Ozenne, Val{\'{e}}ry and
  Blackledge, Martin and Jensen, Malene Ringkj{\o}bing}]{Kragelj2013}
\textbf{\color{eLifeMediumGrey} Kragelj J}, Ozenne V, Blackledge M, Jensen MR.
\newblock {Conformational propensities of intrinsically disordered proteins
  from NMR chemical shifts.}
\newblock Chemphyschem : a European journal of chemical physics and physical
  chemistry.  2013 sep; 14(13):3034--3045.
\newblock \href{10.1002/cphc.201300387}{\doiprefix
  \detokenize{10.1002/cphc.201300387}}.

\bibitem[{Kullback and Leibler(1951)Kullback, S and Leibler, R
  A}]{Kullback1951}
\textbf{\color{eLifeMediumGrey} Kullback S}, Leibler RA.
\newblock {On Information and Sufficiency}.
\newblock The Annals of Mathematical Statistics.  1951 mar; 22(1):79--86.
\newblock \urlprefix\url{https://doi.org/10.1214/aoms/1177729694},
  \href{10.1214/aoms/1177729694}{\doiprefix
  \detokenize{10.1214/aoms/1177729694}}.

\bibitem[{K{\"{u}}mmerer et~al.(2021)K{\"{u}}mmerer, Felix and Orioli, Simone
  and Harding-Larsen, David and Hoffmann, Falk and Gavrilov, Yulian and Teilum,
  Kaare and Lindorff-Larsen, Kresten}]{Kummerer2021}
\textbf{\color{eLifeMediumGrey} K{\"{u}}mmerer F}, Orioli S, Harding-Larsen D,
  Hoffmann F, Gavrilov Y, Teilum K, Lindorff-Larsen K.
\newblock {Fitting Side-Chain NMR Relaxation Data Using Molecular Simulations}.
\newblock Journal of Chemical Theory and Computation.  2021 aug;
  17(8):5262--5275.
\newblock \urlprefix\url{https://doi.org/10.1021/acs.jctc.0c01338},
  \href{10.1021/acs.jctc.0c01338}{\doiprefix
  \detokenize{10.1021/acs.jctc.0c01338}}.

\bibitem[{Larsen et~al.(2020)Larsen, Andreas Haahr and Wang, Yong and Bottaro,
  Sandro and Grudinin, Sergei and Arleth, Lise and Lindorff-Larsen,
  Kresten}]{Larsen2020}
\textbf{\color{eLifeMediumGrey} Larsen AH}, Wang Y, Bottaro S, Grudinin S,
  Arleth L, Lindorff-Larsen K.
\newblock {RESEARCH ARTICLE Combining molecular dynamics simulations with
  small-angle X-ray and neutron scattering data to study multi-domain proteins
  in solution}.
\newblock PLoS Computational Biology.  2020; 16(4):1--29.
\newblock \urlprefix\url{http://dx.doi.org/10.1371/journal.pcbi.1007870},
  \href{10.1371/journal.pcbi.1007870}{\doiprefix
  \detokenize{10.1371/journal.pcbi.1007870}}.

\bibitem[{Lerner et~al.(2021)Lerner, Eitan and Barth, Anders and Hendrix, Jelle
  and Ambrose, Benjamin and Birkedal, Victoria and Blanchard, Scott C and
  B{\"{o}}rner, Richard and {Sung Chung}, Hoi and Cordes, Thorben and Craggs,
  Timothy D and Deniz, Ashok A and Diao, Jiajie and Fei, Jingyi and Gonzalez,
  Ruben L and Gopich, Irina V and Ha, Taekjip and Hanke, Christian A and Haran,
  Gilad and Hatzakis, Nikos S and Hohng, Sungchul and Hong, Seok-Cheol and
  Hugel, Thorsten and Ingargiola, Antonino and Joo, Chirlmin and Kapanidis,
  Achillefs N and Kim, Harold D and Laurence, Ted and Lee, Nam Ki and Lee,
  Tae-Hee and Lemke, Edward A and Margeat, Emmanuel and Michaelis, Jens and
  Michalet, Xavier and Myong, Sua and Nettels, Daniel and Peulen, Thomas-Otavio
  and Ploetz, Evelyn and Razvag, Yair and Robb, Nicole C and Schuler, Benjamin
  and Soleimaninejad, Hamid and Tang, Chun and Vafabakhsh, Reza and Lamb, Don C
  and Seidel, Claus A M and Weiss, Shimon}]{Lerner2021}
\textbf{\color{eLifeMediumGrey} Lerner E}, Barth A, Hendrix J, Ambrose B,
  Birkedal V, Blanchard SC, B{\"{o}}rner R, {Sung Chung} H, Cordes T, Craggs
  TD, Deniz AA, Diao J, Fei J, Gonzalez RL, Gopich IV, Ha T, Hanke CA, Haran G,
  Hatzakis NS, Hohng S, et~al.
\newblock {FRET-based dynamic structural biology: Challenges, perspectives and
  an appeal for open-science practices}.
\newblock eLife.  2021; 10:e60416.
\newblock \urlprefix\url{https://doi.org/10.7554/eLife.60416},
  \href{10.7554/eLife.60416}{\doiprefix \detokenize{10.7554/eLife.60416}}.

\bibitem[{Li and Br{\"{u}}schweiler(2010)Li, Da-Wei and Br{\"{u}}schweiler,
  Rafael}]{Li2010}
\textbf{\color{eLifeMediumGrey} Li DW}, Br{\"{u}}schweiler R.
\newblock {NMR-Based Protein Potentials}.
\newblock Angewandte Chemie International Edition.  2010 sep;
  49(38):6778--6780.
\newblock \urlprefix\url{https://doi.org/10.1002/anie.201001898},
  \href{https://doi.org/10.1002/anie.201001898}{\doiprefix
  \detokenize{https://doi.org/10.1002/anie.201001898}}.

\bibitem[{Lindorff-Larsen et~al.(2005)Lindorff-Larsen, Kresten and Best, Robert
  B and DePristo, Mark A and Dobson, Christopher M and Vendruscolo,
  Michele}]{Lindorff-Larsen2005}
\textbf{\color{eLifeMediumGrey} Lindorff-Larsen K}, Best RB, DePristo MA,
  Dobson CM, Vendruscolo M.
\newblock {Simultaneous determination of protein structure and dynamics}.
\newblock Nature.  2005; 433(7022):128--132.
\newblock \urlprefix\url{https://doi.org/10.1038/nature03199},
  \href{10.1038/nature03199}{\doiprefix \detokenize{10.1038/nature03199}}.

\bibitem[{Lindorff-Larsen and Kragelund(2021)Lindorff-Larsen, Kresten and
  Kragelund, Birthe B}]{Lindorff-Larsen2021}
\textbf{\color{eLifeMediumGrey} Lindorff-Larsen K}, Kragelund BB.
\newblock {On the Potential of Machine Learning to Examine the Relationship
  Between Sequence, Structure, Dynamics and Function of Intrinsically
  Disordered Proteins}.
\newblock Journal of Molecular Biology.  2021; 433(20):167196.
\newblock
  \urlprefix\url{https://www.sciencedirect.com/science/article/pii/S0022283621004290},
  \href{https://doi.org/10.1016/j.jmb.2021.167196}{\doiprefix
  \detokenize{https://doi.org/10.1016/j.jmb.2021.167196}}.

\bibitem[{Lindorff-Larsen et~al.(2012)Lindorff-Larsen, Kresten and Maragakis,
  Paul and Piana, Stefano and Eastwood, Michael P and Dror, Ron O and Shaw,
  David E}]{Lindorff-Larsen2012}
\textbf{\color{eLifeMediumGrey} Lindorff-Larsen K}, Maragakis P, Piana S,
  Eastwood MP, Dror RO, Shaw DE.
\newblock {Systematic Validation of Protein Force Fields against Experimental
  Data}.
\newblock PLOS ONE.  2012 feb; 7(2):e32131.
\newblock \urlprefix\url{https://doi.org/10.1371/journal.pone.0032131}.

\bibitem[{Lipari and Szabo(1982)Lipari, Giovanni and Szabo,
  Attila}]{Lipari1982}
\textbf{\color{eLifeMediumGrey} Lipari G}, Szabo A.
\newblock {Model-free approach to the interpretation of nuclear magnetic
  resonance relaxation in macromolecules. 1. Theory and range of validity}.
\newblock Journal of the American Chemical Society.  1982 aug;
  104(17):4546--4559.
\newblock \urlprefix\url{https://doi.org/10.1021/ja00381a009},
  \href{10.1021/ja00381a009}{\doiprefix \detokenize{10.1021/ja00381a009}}.

\bibitem[{Marion(2013)Marion, Dominique}]{Marion2013}
\textbf{\color{eLifeMediumGrey} Marion D}.
\newblock {An introduction to biological NMR spectroscopy}.
\newblock Molecular \& cellular proteomics : MCP.  2013 nov; 12(11):3006--3025.
\newblock \urlprefix\url{https://pubmed.ncbi.nlm.nih.gov/23831612
  https://www.ncbi.nlm.nih.gov/pmc/articles/PMC3820920/},
  \href{10.1074/mcp.O113.030239}{\doiprefix
  \detokenize{10.1074/mcp.O113.030239}}.

\bibitem[{Marrink et~al.(2007)Marrink, Siewert J. and Risselada, H. Jelger and
  Yefimov, Serge and Tieleman, D. Peter and {De Vries}, Alex H.}]{Marrink2007}
\textbf{\color{eLifeMediumGrey} Marrink SJ}, Risselada HJ, Yefimov S, Tieleman
  DP, {De Vries} AH.
\newblock {The MARTINI force field: Coarse grained model for biomolecular
  simulations}.
\newblock Journal of Physical Chemistry B.  2007; 111(27):7812--7824.
\newblock \href{10.1021/jp071097f}{\doiprefix \detokenize{10.1021/jp071097f}}.

\bibitem[{Martin and Holehouse(2020)Martin, Erik W and Holehouse, Alex
  S}]{Martin2020}
\textbf{\color{eLifeMediumGrey} Martin EW}, Holehouse AS.
\newblock {Intrinsically disordered protein regions and phase separation:
  sequence determinants of assembly or lack thereof}.
\newblock Emerging Topics in Life Sciences.  2020 oct; 4(3):307--329.
\newblock \urlprefix\url{https://doi.org/10.1042/ETLS20190164},
  \href{10.1042/ETLS20190164}{\doiprefix \detokenize{10.1042/ETLS20190164}}.

\bibitem[{Martin et~al.(2021)Martin, Erik W. and Thomasen, F. Emil and
  Milkovic, Nicole M. and Cuneo, Matthew J. and Grace, Christy R. and Nourse,
  Amanda and Lindorff-Larsen, Kresten and Mittag, Tanja}]{Martin2021}
\textbf{\color{eLifeMediumGrey} Martin EW}, Thomasen FE, Milkovic NM, Cuneo MJ,
  Grace CR, Nourse A, Lindorff-Larsen K, Mittag T.
\newblock {Interplay of folded domains and the disordered low-complexity domain
  in mediating hnRNPA1 phase separation}.
\newblock Nucleic Acids Research.  2021; 49(5):2931--2945.
\newblock \href{10.1093/nar/gkab063}{\doiprefix
  \detokenize{10.1093/nar/gkab063}}.

\bibitem[{Metallo(2010)Metallo, Steven J}]{Metallo2010}
\textbf{\color{eLifeMediumGrey} Metallo SJ}.
\newblock {Intrinsically disordered proteins are potential drug targets}.
\newblock Current opinion in chemical biology.  2010 aug; 14(4):481--488.
\newblock \urlprefix\url{https://pubmed.ncbi.nlm.nih.gov/20598937
  https://www.ncbi.nlm.nih.gov/pmc/articles/PMC2918680/},
  \href{10.1016/j.cbpa.2010.06.169}{\doiprefix
  \detokenize{10.1016/j.cbpa.2010.06.169}}.

\bibitem[{Metskas and Rhoades(2020)Metskas, Lauren Ann and Rhoades,
  Elizabeth}]{Metskas2020}
\textbf{\color{eLifeMediumGrey} Metskas LA}, Rhoades E.
\newblock {Single-Molecule FRET of Intrinsically Disordered Proteins}.
\newblock Annual Review of Physical Chemistry.  2020 apr; 71(1):391--414.
\newblock
  \urlprefix\url{https://doi.org/10.1146/annurev-physchem-012420-104917},
  \href{10.1146/annurev-physchem-012420-104917}{\doiprefix
  \detokenize{10.1146/annurev-physchem-012420-104917}}.

\bibitem[{Mioduszewski and Cieplak(2018)Mioduszewski, {\L}ukasz and Cieplak,
  Marek}]{Mioduszewski2018}
\textbf{\color{eLifeMediumGrey} Mioduszewski {\L}}, Cieplak M.
\newblock {Disordered peptide chains in an $\alpha$-C-based coarse-grained
  model}.
\newblock Physical Chemistry Chemical Physics.  2018; 20(28):19057--19070.
\newblock \urlprefix\url{http://dx.doi.org/10.1039/C8CP03309A},
  \href{10.1039/C8CP03309A}{\doiprefix \detokenize{10.1039/C8CP03309A}}.

\bibitem[{Monticelli et~al.(2008)Monticelli, Luca and Kandasamy, Senthil K. and
  Periole, Xavier and Larson, Ronald G. and Tieleman, D. Peter and Marrink,
  Siewert Jan}]{Monticelli2008}
\textbf{\color{eLifeMediumGrey} Monticelli L}, Kandasamy SK, Periole X, Larson
  RG, Tieleman DP, Marrink SJ.
\newblock {The MARTINI coarse-grained force field: Extension to proteins}.
\newblock Journal of Chemical Theory and Computation.  2008; 4(5):819--834.
\newblock \href{10.1021/ct700324x}{\doiprefix \detokenize{10.1021/ct700324x}}.

\bibitem[{Nagy et~al.(2019)Nagy, Gabor and Igaev, Maxim and Jones, Nykola C and
  Hoffmann, S{\o}ren V and Grubm{\"{u}}ller, Helmut}]{Nagy2019}
\textbf{\color{eLifeMediumGrey} Nagy G}, Igaev M, Jones NC, Hoffmann SV,
  Grubm{\"{u}}ller H.
\newblock {SESCA: Predicting Circular Dichroism Spectra from Protein Molecular
  Structures}.
\newblock Journal of Chemical Theory and Computation.  2019 sep;
  15(9):5087--5102.
\newblock \urlprefix\url{https://doi.org/10.1021/acs.jctc.9b00203},
  \href{10.1021/acs.jctc.9b00203}{\doiprefix
  \detokenize{10.1021/acs.jctc.9b00203}}.

\bibitem[{Naudi-Fabra et~al.(2021)Naudi-Fabra, Samuel and Tengo, Maud and
  Jensen, Malene Ringkj{\o}bing and Blackledge, Martin and Milles,
  Sigrid}]{naudi2021quantitative}
\textbf{\color{eLifeMediumGrey} Naudi-Fabra S}, Tengo M, Jensen MR, Blackledge
  M, Milles S.
\newblock Quantitative Description of Intrinsically Disordered Proteins Using
  Single-Molecule FRET, NMR, and SAXS.
\newblock Journal of the American Chemical Society.  2021; .

\bibitem[{No{\'{e}} et~al.(2020)No{\'{e}}, Frank and Tkatchenko, Alexandre and
  M{\"{u}}ller, Klaus-Robert and Clementi, Cecilia}]{Noe2020}
\textbf{\color{eLifeMediumGrey} No{\'{e}} F}, Tkatchenko A, M{\"{u}}ller KR,
  Clementi C.
\newblock {Machine Learning for Molecular Simulation}.
\newblock Annual Review of Physical Chemistry.  2020 apr; 71(1):361--390.
\newblock
  \urlprefix\url{https://doi.org/10.1146/annurev-physchem-042018-052331},
  \href{10.1146/annurev-physchem-042018-052331}{\doiprefix
  \detokenize{10.1146/annurev-physchem-042018-052331}}.

\bibitem[{Norgaard et~al.(2008)Norgaard, Anders B and Ferkinghoff-Borg, Jesper
  and Lindorff-Larsen, Kresten}]{Norgaard2008}
\textbf{\color{eLifeMediumGrey} Norgaard AB}, Ferkinghoff-Borg J,
  Lindorff-Larsen K.
\newblock {Experimental parameterization of an energy function for the
  simulation of unfolded proteins}.
\newblock Biophysical journal.  2008 jan; 94(1):182--192.
\newblock \urlprefix\url{https://pubmed.ncbi.nlm.nih.gov/17827232
  https://www.ncbi.nlm.nih.gov/pmc/articles/PMC2134871/},
  \href{10.1529/biophysj.107.108241}{\doiprefix
  \detokenize{10.1529/biophysj.107.108241}}.

\bibitem[{Oates et~al.(2013)Oates, Matt E and Romero, Pedro and Ishida, Takashi
  and Ghalwash, Mohamed and Mizianty, Marcin J and Xue, Bin and
  Doszt{\'{a}}nyi, Zsuzsanna and Uversky, Vladimir N and Obradovic, Zoran and
  Kurgan, Lukasz and Dunker, A Keith and Gough, Julian}]{Oates2013}
\textbf{\color{eLifeMediumGrey} Oates ME}, Romero P, Ishida T, Ghalwash M,
  Mizianty MJ, Xue B, Doszt{\'{a}}nyi Z, Uversky VN, Obradovic Z, Kurgan L,
  Dunker AK, Gough J.
\newblock {D2P2: database of disordered protein predictions}.
\newblock Nucleic Acids Research.  2013 jan; 41(D1):D508--D516.
\newblock \urlprefix\url{https://doi.org/10.1093/nar/gks1226},
  \href{10.1093/nar/gks1226}{\doiprefix \detokenize{10.1093/nar/gks1226}}.

\bibitem[{Olsson et~al.(2013)Olsson, Simon and Frellsen, Jes and Boomsma,
  Wouter and Mardia, Kanti V and Hamelryck, Thomas}]{Olsson2013}
\textbf{\color{eLifeMediumGrey} Olsson S}, Frellsen J, Boomsma W, Mardia KV,
  Hamelryck T.
\newblock {Inference of Structure Ensembles of Flexible Biomolecules from
  Sparse, Averaged Data}.
\newblock PLOS ONE.  2013 nov; 8(11):e79439.
\newblock \urlprefix\url{https://doi.org/10.1371/journal.pone.0079439}.

\bibitem[{Orioli et~al.(2020)Orioli, Simone and Larsen, Andreas Haahr and
  Bottaro, Sandro and Lindorff-Larsen, Kresten}]{Orioli2020}
\textbf{\color{eLifeMediumGrey} Orioli S}, Larsen AH, Bottaro S,
  Lindorff-Larsen K.
\newblock {Chapter Three - How to learn from inconsistencies: Integrating
  molecular simulations with experimental data}.
\newblock In: Strodel B, Barz BBTPiMB, Science T, editors. \emph{Computational
  Approaches for Understanding Dynamical Systems: Protein Folding and
  Assembly}, vol. 170 Academic Press; 2020.p. 123--176.
\newblock
  \urlprefix\url{http://www.sciencedirect.com/science/article/pii/S1877117319302121},
  \href{https://doi.org/10.1016/bs.pmbts.2019.12.006}{\doiprefix
  \detokenize{https://doi.org/10.1016/bs.pmbts.2019.12.006}}.

\bibitem[{Ozenne et~al.(2012{\natexlab{a}})Ozenne, Val{\'{e}}ry and Bauer,
  Fr{\'{e}}d{\'{e}}ric and Salmon, Lo{\"{i}}c and Huang, Jie Rong and Jensen,
  Malene Ringkj{\o}bing and Segard, St{\'{e}}phane and Bernad{\'{o}}, Pau and
  Charavay, C{\'{e}}line and Blackledge, Martin}]{Ozenne2012}
\textbf{\color{eLifeMediumGrey} Ozenne V}, Bauer F, Salmon L, Huang JR, Jensen
  MR, Segard S, Bernad{\'{o}} P, Charavay C, Blackledge M.
\newblock {Flexible-meccano: A tool for the generation of explicit ensemble
  descriptions of intrinsically disordered proteins and their associated
  experimental observables}.
\newblock Bioinformatics.  2012; 28(11):1463--1470.
\newblock \href{10.1093/bioinformatics/bts172}{\doiprefix
  \detokenize{10.1093/bioinformatics/bts172}}.

\bibitem[{Ozenne et~al.(2012{\natexlab{b}})Ozenne, Valéry and Schneider,
  Robert and Yao, Mingxi and Huang, Jie-rong and Salmon, Lo{\"\i}c and
  Zweckstetter, Markus and Jensen, Malene Ringkj{\o}bing and Blackledge,
  Martin}]{ozenne2012mapping}
\textbf{\color{eLifeMediumGrey} Ozenne V}, Schneider R, Yao M, Huang Jr, Salmon
  L, Zweckstetter M, Jensen MR, Blackledge M.
\newblock Mapping the potential energy landscape of intrinsically disordered
  proteins at amino acid resolution.
\newblock Journal of the American Chemical Society.  2012;
  134(36):15138--15148.

\bibitem[{Palazzesi et~al.(2015)Palazzesi, Ferruccio and Prakash, Meher K and
  Bonomi, Massimiliano and Barducci, Alessandro}]{Palazzesi2015}
\textbf{\color{eLifeMediumGrey} Palazzesi F}, Prakash MK, Bonomi M, Barducci A.
\newblock {Accuracy of Current All-Atom Force-Fields in Modeling Protein
  Disordered States}.
\newblock Journal of Chemical Theory and Computation.  2015 jan; 11(1):2--7.
\newblock \urlprefix\url{https://doi.org/10.1021/ct500718s},
  \href{10.1021/ct500718s}{\doiprefix \detokenize{10.1021/ct500718s}}.

\bibitem[{Pannier et~al.(2000)Pannier, M and Veit, S and Godt, A and Jeschke, G
  and Spiess, H W}]{Pannier2000}
\textbf{\color{eLifeMediumGrey} Pannier M}, Veit S, Godt A, Jeschke G, Spiess
  HW.
\newblock {Dead-Time Free Measurement of Dipole–Dipole Interactions between
  Electron Spins}.
\newblock Journal of Magnetic Resonance.  2000; 142(2):331--340.
\newblock
  \urlprefix\url{https://www.sciencedirect.com/science/article/pii/S1090780799919444},
  \href{https://doi.org/10.1006/jmre.1999.1944}{\doiprefix
  \detokenize{https://doi.org/10.1006/jmre.1999.1944}}.

\bibitem[{Peran and Mittag(2020)Peran, Ivan and Mittag, Tanja}]{Peran2020}
\textbf{\color{eLifeMediumGrey} Peran I}, Mittag T.
\newblock {Molecular structure in biomolecular condensates}.
\newblock Current Opinion in Structural Biology.  2020; 60:17--26.
\newblock
  \urlprefix\url{https://www.sciencedirect.com/science/article/pii/S0959440X19301083},
  \href{https://doi.org/10.1016/j.sbi.2019.09.007}{\doiprefix
  \detokenize{https://doi.org/10.1016/j.sbi.2019.09.007}}.

\bibitem[{Pesce and Lindorff-Larsen(2021)Pesce, Francesco and Lindorff-Larsen,
  Kresten}]{Pesce2021}
\textbf{\color{eLifeMediumGrey} Pesce F}, Lindorff-Larsen K.
\newblock {Refining conformational ensembles of flexible proteins against
  small-angle X-ray scattering data}.
\newblock bioRxiv.  2021 jan; p. 2021.05.29.446281.
\newblock
  \urlprefix\url{http://biorxiv.org/content/early/2021/09/09/2021.05.29.446281.abstract},
  \href{10.1101/2021.05.29.446281}{\doiprefix
  \detokenize{10.1101/2021.05.29.446281}}.

\bibitem[{Peter et~al.(2001)Peter, Christine and Daura, Xavier and van
  Gunsteren, Wilfred F}]{Peter2001}
\textbf{\color{eLifeMediumGrey} Peter C}, Daura X, van Gunsteren WF.
\newblock {Calculation of NMR-relaxation parameters for flexible molecules from
  molecular dynamics simulations}.
\newblock Journal of Biomolecular NMR.  2001; 20(4):297--310.
\newblock \urlprefix\url{https://doi.org/10.1023/A:1011241030461},
  \href{10.1023/A:1011241030461}{\doiprefix
  \detokenize{10.1023/A:1011241030461}}.

\bibitem[{Petoukhov et~al.(2012)Petoukhov, Maxim V and Franke, Daniel and
  Shkumatov, Alexander V and Tria, Giancarlo and Kikhney, Alexey G and Gajda,
  Michal and Gorba, Christian and Mertens, Haydyn D T and Konarev, Petr V and
  Svergun, Dmitri I}]{Petoukhov:fs5015}
\textbf{\color{eLifeMediumGrey} Petoukhov MV}, Franke D, Shkumatov AV, Tria G,
  Kikhney AG, Gajda M, Gorba C, Mertens HDT, Konarev PV, Svergun DI.
\newblock {New developments in the {\it ATSAS} program package for small-angle
  scattering data analysis}.
\newblock Journal of Applied Crystallography.  2012 apr; 45(2):342--350.
\newblock \urlprefix\url{https://doi.org/10.1107/S0021889812007662},
  \href{10.1107/S0021889812007662}{\doiprefix
  \detokenize{10.1107/S0021889812007662}}.

\bibitem[{Petoukhov and Svergun(2005)Petoukhov, Maxim V. and Svergun, Dmitri
  I.}]{Petoukhov2005}
\textbf{\color{eLifeMediumGrey} Petoukhov MV}, Svergun DI.
\newblock {Global rigid body modeling of macromolecular complexes against
  small-angle scattering data}.
\newblock Biophysical Journal.  2005; 89(2):1237--1250.
\newblock \href{10.1529/biophysj.105.064154}{\doiprefix
  \detokenize{10.1529/biophysj.105.064154}}.

\bibitem[{Piana et~al.(2015)Piana, Stefano and Donchev, Alexander G and
  Robustelli, Paul and Shaw, David E}]{Piana2015}
\textbf{\color{eLifeMediumGrey} Piana S}, Donchev AG, Robustelli P, Shaw DE.
\newblock {Water Dispersion Interactions Strongly Influence Simulated
  Structural Properties of Disordered Protein States}.
\newblock The Journal of Physical Chemistry B.  2015 apr; 119(16):5113--5123.
\newblock \urlprefix\url{https://doi.org/10.1021/jp508971m},
  \href{10.1021/jp508971m}{\doiprefix \detokenize{10.1021/jp508971m}}.

\bibitem[{Pietrek et~al.(2020)Pietrek, Lisa M and Stelzl, Lukas S and Hummer,
  Gerhard}]{Pietrek2020}
\textbf{\color{eLifeMediumGrey} Pietrek LM}, Stelzl LS, Hummer G.
\newblock {Hierarchical Ensembles of Intrinsically Disordered Proteins at
  Atomic Resolution in Molecular Dynamics Simulations}.
\newblock Journal of Chemical Theory and Computation.  2020 jan;
  16(1):725--737.
\newblock \urlprefix\url{https://doi.org/10.1021/acs.jctc.9b00809},
  \href{10.1021/acs.jctc.9b00809}{\doiprefix
  \detokenize{10.1021/acs.jctc.9b00809}}.

\bibitem[{Pitera and Chodera(2012)Pitera, Jed W and Chodera, John
  D}]{Pitera2012}
\textbf{\color{eLifeMediumGrey} Pitera JW}, Chodera JD.
\newblock {On the Use of Experimental Observations to Bias Simulated
  Ensembles}.
\newblock Journal of Chemical Theory and Computation.  2012 oct;
  8(10):3445--3451.
\newblock \urlprefix\url{https://doi.org/10.1021/ct300112v},
  \href{10.1021/ct300112v}{\doiprefix \detokenize{10.1021/ct300112v}}.

\bibitem[{Polyhach et~al.(2011)Polyhach, Yevhen and Bordignon, Enrica and
  Jeschke, Gunnar}]{Polyhach2011}
\textbf{\color{eLifeMediumGrey} Polyhach Y}, Bordignon E, Jeschke G.
\newblock {Rotamer libraries of spin labelled cysteines for protein studies}.
\newblock Phys Chem Chem Phys.  2011; 13(6):2356--2366.
\newblock \urlprefix\url{https://doi.org/10.1039/c0cp01865a}.

\bibitem[{Pond et~al.(2020)Pond, Matthew P and Eells, Rebecca and Treece,
  Bradley W and Heinrich, Frank and L{\"{o}}sche, Mathias and Roux,
  Beno{\^{i}}t}]{Pond2020}
\textbf{\color{eLifeMediumGrey} Pond MP}, Eells R, Treece BW, Heinrich F,
  L{\"{o}}sche M, Roux B.
\newblock {Membrane Anchoring of Hck Kinase via the Intrinsically Disordered
  SH4-U and Length Scale Associated with Subcellular Localization}.
\newblock Journal of Molecular Biology.  2020; 432(9):2985--2997.
\newblock
  \urlprefix\url{https://www.sciencedirect.com/science/article/pii/S0022283619307211},
  \href{https://doi.org/10.1016/j.jmb.2019.11.024}{\doiprefix
  \detokenize{https://doi.org/10.1016/j.jmb.2019.11.024}}.

\bibitem[{Rangan et~al.(2018)Rangan, Ramya and Bonomi, Massimiliano and Heller,
  Gabriella T and Cesari, Andrea and Bussi, Giovanni and Vendruscolo,
  Michele}]{rangan2018determination}
\textbf{\color{eLifeMediumGrey} Rangan R}, Bonomi M, Heller GT, Cesari A, Bussi
  G, Vendruscolo M.
\newblock Determination of structural ensembles of proteins: restraining vs
  reweighting.
\newblock Journal of chemical theory and computation.  2018; 14(12):6632--6641.

\bibitem[{Rauscher et~al.(2015)Rauscher, Sarah and Gapsys, Vytautas and Gajda,
  Michal J and Zweckstetter, Markus and de Groot, Bert L and Grubm{\"{u}}ller,
  Helmut}]{Rauscher2015}
\textbf{\color{eLifeMediumGrey} Rauscher S}, Gapsys V, Gajda MJ, Zweckstetter
  M, de~Groot BL, Grubm{\"{u}}ller H.
\newblock {Structural Ensembles of Intrinsically Disordered Proteins Depend
  Strongly on Force Field: A Comparison to Experiment}.
\newblock Journal of Chemical Theory and Computation.  2015 nov;
  11(11):5513--5524.
\newblock \urlprefix\url{https://doi.org/10.1021/acs.jctc.5b00736},
  \href{10.1021/acs.jctc.5b00736}{\doiprefix
  \detokenize{10.1021/acs.jctc.5b00736}}.

\bibitem[{Regy et~al.(2021)Regy, Roshan Mammen and Thompson, Jacob and Kim,
  Young C and Mittal, Jeetain}]{Regy2021}
\textbf{\color{eLifeMediumGrey} Regy RM}, Thompson J, Kim YC, Mittal J.
\newblock {Improved coarse-grained model for studying sequence dependent phase
  separation of disordered proteins.}
\newblock Protein science : a publication of the Protein Society.  2021 jul;
  30(7):1371--1379.
\newblock \href{10.1002/pro.4094}{\doiprefix \detokenize{10.1002/pro.4094}}.

\bibitem[{Reichel et~al.(2018)Reichel, Katrin and Stelzl, Lukas S and
  K{\"{o}}finger, J{\"{u}}rgen and Hummer, Gerhard}]{Reichel2018}
\textbf{\color{eLifeMediumGrey} Reichel K}, Stelzl LS, K{\"{o}}finger J, Hummer
  G.
\newblock {Precision DEER Distances from Spin-Label Ensemble Refinement}.
\newblock The Journal of Physical Chemistry Letters.  2018 oct;
  9(19):5748--5752.
\newblock \urlprefix\url{https://doi.org/10.1021/acs.jpclett.8b02439},
  \href{10.1021/acs.jpclett.8b02439}{\doiprefix
  \detokenize{10.1021/acs.jpclett.8b02439}}.

\bibitem[{Rennella and Brutscher(2013)Rennella, Enrico and Brutscher,
  Bernhard}]{Rennella2013}
\textbf{\color{eLifeMediumGrey} Rennella E}, Brutscher B.
\newblock {Fast Real-Time NMR Methods for Characterizing Short-Lived Molecular
  States}.
\newblock ChemPhysChem.  2013 sep; 14(13):3059--3070.
\newblock \urlprefix\url{https://doi.org/10.1002/cphc.201300339},
  \href{https://doi.org/10.1002/cphc.201300339}{\doiprefix
  \detokenize{https://doi.org/10.1002/cphc.201300339}}.

\bibitem[{Reppert et~al.(2016)Reppert, Mike and Roy, Anish R and Tempkin,
  Jeremy O B and Dinner, Aaron R and Tokmakoff, Andrei}]{Reppert2016}
\textbf{\color{eLifeMediumGrey} Reppert M}, Roy AR, Tempkin JOB, Dinner AR,
  Tokmakoff A.
\newblock {Refining Disordered Peptide Ensembles with Computational Amide I
  Spectroscopy: Application to Elastin-Like Peptides}.
\newblock The Journal of Physical Chemistry B.  2016 nov; 120(44):11395--11404.
\newblock \urlprefix\url{https://doi.org/10.1021/acs.jpcb.6b08678},
  \href{10.1021/acs.jpcb.6b08678}{\doiprefix
  \detokenize{10.1021/acs.jpcb.6b08678}}.

\bibitem[{Reppert and Tokmakoff(2013)Reppert, Mike and Tokmakoff,
  Andrei}]{Reppert2013}
\textbf{\color{eLifeMediumGrey} Reppert M}, Tokmakoff A.
\newblock {Electrostatic frequency shifts in amide I vibrational spectra:
  Direct parameterization against experiment}.
\newblock The Journal of Chemical Physics.  2013 apr; 138(13):134116.
\newblock \urlprefix\url{https://doi.org/10.1063/1.4798938},
  \href{10.1063/1.4798938}{\doiprefix \detokenize{10.1063/1.4798938}}.

\bibitem[{Robustelli et~al.(2018)Robustelli, Paul and Piana, Stefano and Shaw,
  David E}]{Robustelli2018}
\textbf{\color{eLifeMediumGrey} Robustelli P}, Piana S, Shaw DE.
\newblock {Developing a molecular dynamics force field for both folded and
  disordered protein states}.
\newblock Proceedings of the National Academy of Sciences.  2018 may;
  115(21):E4758 LP -- E4766.
\newblock \urlprefix\url{http://www.pnas.org/content/115/21/E4758.abstract},
  \href{10.1073/pnas.1800690115}{\doiprefix
  \detokenize{10.1073/pnas.1800690115}}.

\bibitem[{Roux and Weare(2013)Roux, Beno{\^{i}}t and Weare,
  Jonathan}]{Roux2013}
\textbf{\color{eLifeMediumGrey} Roux B}, Weare J.
\newblock {On the statistical equivalence of restrained-ensemble simulations
  with the maximum entropy method}.
\newblock The Journal of Chemical Physics.  2013 feb; 138(8):84107.
\newblock \urlprefix\url{https://doi.org/10.1063/1.4792208},
  \href{10.1063/1.4792208}{\doiprefix \detokenize{10.1063/1.4792208}}.

\bibitem[{Roy et~al.(2016)Roy, Amitava and Hua, Duy P and Post, Carol
  Beth}]{Roy2016}
\textbf{\color{eLifeMediumGrey} Roy A}, Hua DP, Post CB.
\newblock {Analysis of Multidomain Protein Dynamics}.
\newblock Journal of Chemical Theory and Computation.  2016 jan;
  12(1):274--280.
\newblock \urlprefix\url{https://doi.org/10.1021/acs.jctc.5b00796},
  \href{10.1021/acs.jctc.5b00796}{\doiprefix
  \detokenize{10.1021/acs.jctc.5b00796}}.

\bibitem[{R{\'{o}}{\.{z}}ycki et~al.(2011)R{\'{o}}{\.{z}}ycki, Bartosz and Kim,
  Young C and Hummer, Gerhard}]{Rozycki2011}
\textbf{\color{eLifeMediumGrey} R{\'{o}}{\. {z}}ycki B}, Kim YC, Hummer G.
\newblock {SAXS Ensemble Refinement of ESCRT-III CHMP3 Conformational
  Transitions}.
\newblock Structure.  2011; 19(1):109--116.
\newblock
  \urlprefix\url{https://www.sciencedirect.com/science/article/pii/S0969212610003953},
  \href{https://doi.org/10.1016/j.str.2010.10.006}{\doiprefix
  \detokenize{https://doi.org/10.1016/j.str.2010.10.006}}.

\bibitem[{Rutter et~al.(2015)Rutter, Gil O and Brown, Aaron H and Quigley,
  David and Walsh, Tiffany R and Allen, Michael P}]{Rutter2015}
\textbf{\color{eLifeMediumGrey} Rutter GO}, Brown AH, Quigley D, Walsh TR,
  Allen MP.
\newblock {Testing the transferability of a coarse-grained model to
  intrinsically disordered proteins}.
\newblock Physical Chemistry Chemical Physics.  2015; 17(47):31741--31749.
\newblock \urlprefix\url{http://dx.doi.org/10.1039/C5CP05652G},
  \href{10.1039/C5CP05652G}{\doiprefix \detokenize{10.1039/C5CP05652G}}.

\bibitem[{Saad et~al.(2021)Saad, Dana and Paissoni, Cristina and
  Chaves-Sanjuan, Antonio and Nardini, Marco and Mantovani, Roberto and
  Gnesutta, Nerina and Camilloni, Carlo}]{Saad2021}
\textbf{\color{eLifeMediumGrey} Saad D}, Paissoni C, Chaves-Sanjuan A, Nardini
  M, Mantovani R, Gnesutta N, Camilloni C.
\newblock {High Conformational Flexibility of the E2F1/DP1/DNA Complex}.
\newblock Journal of Molecular Biology.  2021; 433(18):167119.
\newblock
  \urlprefix\url{https://www.sciencedirect.com/science/article/pii/S0022283621003430},
  \href{https://doi.org/10.1016/j.jmb.2021.167119}{\doiprefix
  \detokenize{https://doi.org/10.1016/j.jmb.2021.167119}}.

\bibitem[{Salmon et~al.(2010)Salmon, Lo{\"{i}}c and Nodet, Gabrielle and
  Ozenne, Val{\'{e}}ry and Yin, Guowei and Jensen, Malene Ringkj{\o}bing and
  Zweckstetter, Markus and Blackledge, Martin}]{Salmon2010}
\textbf{\color{eLifeMediumGrey} Salmon L}, Nodet G, Ozenne V, Yin G, Jensen MR,
  Zweckstetter M, Blackledge M.
\newblock {NMR Characterization of Long-Range Order in Intrinsically Disordered
  Proteins}.
\newblock Journal of the American Chemical Society.  2010 jun;
  132(24):8407--8418.
\newblock \urlprefix\url{https://doi.org/10.1021/ja101645g},
  \href{10.1021/ja101645g}{\doiprefix \detokenize{10.1021/ja101645g}}.

\bibitem[{Salvi et~al.(2016)Salvi, Nicola and Abyzov, Anton and Blackledge,
  Martin}]{Salvi2016}
\textbf{\color{eLifeMediumGrey} Salvi N}, Abyzov A, Blackledge M.
\newblock {Multi-Timescale Dynamics in Intrinsically Disordered Proteins from
  NMR Relaxation and Molecular Simulation}.
\newblock The Journal of Physical Chemistry Letters.  2016 jul;
  7(13):2483--2489.
\newblock \urlprefix\url{https://doi.org/10.1021/acs.jpclett.6b00885},
  \href{10.1021/acs.jpclett.6b00885}{\doiprefix
  \detokenize{10.1021/acs.jpclett.6b00885}}.

\bibitem[{Schneidman-Duhovny et~al.(2010)Schneidman-Duhovny, Dina and Hammel,
  Michal and Sali, Andrej}]{Schneidman-Duhovny2010}
\textbf{\color{eLifeMediumGrey} Schneidman-Duhovny D}, Hammel M, Sali A.
\newblock {FoXS: a web server for rapid computation and fitting of SAXS
  profiles.}
\newblock Nucleic acids research.  2010 jul; 38(Web Server issue):W540--4.
\newblock \href{10.1093/nar/gkq461}{\doiprefix
  \detokenize{10.1093/nar/gkq461}}.

\bibitem[{Schneidman-Duhovny et~al.(2013)Schneidman-Duhovny, Dina and Hammel,
  Michal and Tainer, John A and Sali, Andrej}]{Schneidman-Duhovny2013}
\textbf{\color{eLifeMediumGrey} Schneidman-Duhovny D}, Hammel M, Tainer JA,
  Sali A.
\newblock {Accurate SAXS profile computation and its assessment by contrast
  variation experiments.}
\newblock Biophysical journal.  2013 aug; 105(4):962--974.
\newblock \href{10.1016/j.bpj.2013.07.020}{\doiprefix
  \detokenize{10.1016/j.bpj.2013.07.020}}.

\bibitem[{Shannon(1948)Shannon, C E}]{Shannon1948}
\textbf{\color{eLifeMediumGrey} Shannon CE}.
\newblock {A Mathematical Theory of Communication}.
\newblock Bell System Technical Journal.  1948 jul; 27(3):379--423.
\newblock \urlprefix\url{https://doi.org/10.1002/j.1538-7305.1948.tb01338.x},
  \href{https://doi.org/10.1002/j.1538-7305.1948.tb01338.x}{\doiprefix
  \detokenize{https://doi.org/10.1002/j.1538-7305.1948.tb01338.x}}.

\bibitem[{Shen and Bax(2010)Shen, Yang and Bax, Ad}]{Shen2010}
\textbf{\color{eLifeMediumGrey} Shen Y}, Bax A.
\newblock {SPARTA+: a modest improvement in empirical NMR chemical shift
  prediction by means of an artificial neural network}.
\newblock Journal of Biomolecular NMR.  2010; 48(1):13--22.
\newblock \urlprefix\url{https://doi.org/10.1007/s10858-010-9433-9},
  \href{10.1007/s10858-010-9433-9}{\doiprefix
  \detokenize{10.1007/s10858-010-9433-9}}.

\bibitem[{Shen et~al.(2008)Shen, Yang and Lange, Oliver and Delaglio, Frank and
  Rossi, Paolo and Aramini, James M and Liu, Gaohua and Eletsky, Alexander and
  Wu, Yibing and Singarapu, Kiran K and Lemak, Alexander and
  others}]{shen2008consistent}
\textbf{\color{eLifeMediumGrey} Shen Y}, Lange O, Delaglio F, Rossi P, Aramini
  JM, Liu G, Eletsky A, Wu Y, Singarapu KK, Lemak A, et~al.
\newblock Consistent blind protein structure generation from NMR chemical shift
  data.
\newblock Proceedings of the National Academy of Sciences.  2008;
  105(12):4685--4690.

\bibitem[{Sindbert et~al.(2011)Sindbert, Simon and Kalinin, Stanislav and
  Nguyen, Hien and Kienzler, Andrea and Clima, Lilia and Bannwarth, Willi and
  Appel, Bettina and M{\"{u}}ller, Sabine and Seidel, Claus A M}]{Sindbert2011}
\textbf{\color{eLifeMediumGrey} Sindbert S}, Kalinin S, Nguyen H, Kienzler A,
  Clima L, Bannwarth W, Appel B, M{\"{u}}ller S, Seidel CAM.
\newblock {Accurate Distance Determination of Nucleic Acids via F{\"{o}}rster
  Resonance Energy Transfer: Implications of Dye Linker Length and Rigidity}.
\newblock Journal of the American Chemical Society.  2011 mar;
  133(8):2463--2480.
\newblock \urlprefix\url{https://doi.org/10.1021/ja105725e},
  \href{10.1021/ja105725e}{\doiprefix \detokenize{10.1021/ja105725e}}.

\bibitem[{Smith et~al.(2020)Smith, Colin A and Mazur, Adam and Rout, Ashok K
  and Becker, Stefan and Lee, Donghan and de Groot, Bert L and Griesinger,
  Christian}]{Smith2020}
\textbf{\color{eLifeMediumGrey} Smith CA}, Mazur A, Rout AK, Becker S, Lee D,
  de~Groot BL, Griesinger C.
\newblock {Enhancing NMR derived ensembles with kinetics on multiple
  timescales}.
\newblock Journal of Biomolecular NMR.  2020; 74(1):27--43.
\newblock \urlprefix\url{https://doi.org/10.1007/s10858-019-00288-8},
  \href{10.1007/s10858-019-00288-8}{\doiprefix
  \detokenize{10.1007/s10858-019-00288-8}}.

\bibitem[{Souza et~al.(2021)Souza, Paulo C.T. and Alessandri, Riccardo and
  Barnoud, Jonathan and Thallmair, Sebastian and Faustino, Ignacio and
  Gr{\"{u}}newald, Fabian and Patmanidis, Ilias and Abdizadeh, Haleh and
  Bruininks, Bart M.H. and Wassenaar, Tsjerk A. and Kroon, Peter C. and Melcr,
  Josef and Nieto, Vincent and Corradi, Valentina and Khan, Hanif M. and
  Doma{\'{n}}ski, Jan and Javanainen, Matti and Martinez-Seara, Hector and
  Reuter, Nathalie and Best, Robert B. and Vattulainen, Ilpo and Monticelli,
  Luca and Periole, Xavier and Tieleman, D. Peter and de Vries, Alex H. and
  Marrink, Siewert J.}]{Souza2021}
\textbf{\color{eLifeMediumGrey} Souza PCT}, Alessandri R, Barnoud J, Thallmair
  S, Faustino I, Gr{\"{u}}newald F, Patmanidis I, Abdizadeh H, Bruininks BMH,
  Wassenaar TA, Kroon PC, Melcr J, Nieto V, Corradi V, Khan HM, Doma{\'{n}}ski
  J, Javanainen M, Martinez-Seara H, Reuter N, Best RB, et~al.
\newblock {Martini 3: a general purpose force field for coarse-grained
  molecular dynamics}.
\newblock Nature Methods.  2021; 18(4):382--388.
\newblock \href{10.1038/s41592-021-01098-3}{\doiprefix
  \detokenize{10.1038/s41592-021-01098-3}}.

\bibitem[{Steinhoff and Hubbell(1996)Steinhoff, H J and Hubbell, W
  L}]{Steinhoff1996}
\textbf{\color{eLifeMediumGrey} Steinhoff HJ}, Hubbell WL.
\newblock {Calculation of electron paramagnetic resonance spectra from Brownian
  dynamics trajectories: application to nitroxide side chains in proteins}.
\newblock Biophysical Journal.  1996; 71(4):2201--2212.
\newblock
  \urlprefix\url{https://www.sciencedirect.com/science/article/pii/S0006349596794213},
  \href{https://doi.org/10.1016/S0006-3495(96)79421-3}{\doiprefix
  \detokenize{https://doi.org/10.1016/S0006-3495(96)79421-3}}.

\bibitem[{Stelzl et~al.(2021)Stelzl, Lukas S and Pietrek, Lisa M and Holla,
  Andrea and Oroz, Javier S and Sikora, Mateusz and K{\"o}finger, J{\"u}rgen
  and Schuler, Benjamin and Zweckstetter, Markus and Hummer,
  Gerhard}]{stelzl2021global}
\textbf{\color{eLifeMediumGrey} Stelzl LS}, Pietrek LM, Holla A, Oroz JS,
  Sikora M, K{\"o}finger J, Schuler B, Zweckstetter M, Hummer G.
\newblock Global Structure of the Intrinsically Disordered Protein Tau Emerges
  from its Local Structure.
\newblock bioRxiv.  2021; .

\bibitem[{Svergun et~al.(1995)Svergun, D and Barberato, C and Koch, M H
  J}]{Svergun1995}
\textbf{\color{eLifeMediumGrey} Svergun D}, Barberato C, Koch MHJ.
\newblock {CRYSOL– a Program to Evaluate X-ray Solution Scattering of
  Biological Macromolecules from Atomic Coordinates}.
\newblock Journal of Applied Crystallography.  1995 dec; 28(6):768--773.
\newblock \urlprefix\url{https://doi.org/10.1107/S0021889895007047},
  \href{https://doi.org/10.1107/S0021889895007047}{\doiprefix
  \detokenize{https://doi.org/10.1107/S0021889895007047}}.

\bibitem[{Teilum et~al.(2009)Teilum, Kaare and Olsen, Johan G. and Kragelund,
  Birthe B.}]{Teilum2009}
\textbf{\color{eLifeMediumGrey} Teilum K}, Olsen JG, Kragelund BB.
\newblock {Functional aspects of protein flexibility}.
\newblock Cellular and Molecular Life Sciences.  2009; 66(14):2231--2247.
\newblock \href{10.1007/s00018-009-0014-6}{\doiprefix
  \detokenize{10.1007/s00018-009-0014-6}}.

\bibitem[{Tesei et~al.(2021{\natexlab{a}})Tesei, Giulio and Martins, Jo{\~{a}}o
  M and Kunze, Micha B A and Wang, Yong and Crehuet, Ramon and Lindorff-Larsen,
  Kresten}]{Tesei2021}
\textbf{\color{eLifeMediumGrey} Tesei G}, Martins JM, Kunze MBA, Wang Y,
  Crehuet R, Lindorff-Larsen K.
\newblock {DEER-PREdict: Software for efficient calculation of spin-labeling
  EPR and NMR data from conformational ensembles}.
\newblock {PLOS} Computational Biology.  2021 jan; 17(1):e1008551.
\newblock \urlprefix\url{https://doi.org/10.1371/journal.pcbi.1008551},
  \href{10.1371/journal.pcbi.1008551}{\doiprefix
  \detokenize{10.1371/journal.pcbi.1008551}}.

\bibitem[{Tesei et~al.(2021{\natexlab{b}})Tesei, Giulio and Schulze, Thea K and
  Crehuet, Ramon and Lindorff-Larsen, Kresten}]{Tesei2021a}
\textbf{\color{eLifeMediumGrey} Tesei G}, Schulze TK, Crehuet R,
  Lindorff-Larsen K.
\newblock {Accurate model of liquid-liquid phase behaviour of
  intrinsically-disordered proteins from optimization of single-chain
  properties}.
\newblock bioRxiv.  2021 jan; p. 2021.06.23.449550.
\newblock
  \urlprefix\url{http://biorxiv.org/content/early/2021/09/10/2021.06.23.449550.abstract},
  \href{10.1101/2021.06.23.449550}{\doiprefix
  \detokenize{10.1101/2021.06.23.449550}}.

\bibitem[{Thomasen et~al.(2021)Thomasen, F Emil and Pesce, Francesco and
  Roesgaard, Mette Ahrensback and Tesei, Giulio and Lindorff-Larsen,
  Kresten}]{Thomasen2021}
\textbf{\color{eLifeMediumGrey} Thomasen FE}, Pesce F, Roesgaard MA, Tesei G,
  Lindorff-Larsen K.
\newblock {Improving the global dimensions of intrinsically disordered proteins
  in Martini 3}.
\newblock bioRxiv.  2021 jan; p. 2021.10.01.462803.
\newblock
  \urlprefix\url{http://biorxiv.org/content/early/2021/10/01/2021.10.01.462803.abstract},
  \href{10.1101/2021.10.01.462803}{\doiprefix
  \detokenize{10.1101/2021.10.01.462803}}.

\bibitem[{Tombolato et~al.(2006{\natexlab{a}})Tombolato, Fabio and Ferrarini,
  Alberta and Freed, Jack H}]{Tombolato2006a}
\textbf{\color{eLifeMediumGrey} Tombolato F}, Ferrarini A, Freed JH.
\newblock {Dynamics of the Nitroxide Side Chain in Spin-Labeled Proteins}.
\newblock The Journal of Physical Chemistry B.  2006 dec; 110(51):26248--26259.
\newblock \urlprefix\url{https://doi.org/10.1021/jp0629487},
  \href{10.1021/jp0629487}{\doiprefix \detokenize{10.1021/jp0629487}}.

\bibitem[{Tombolato et~al.(2006{\natexlab{b}})Tombolato, Fabio and Ferrarini,
  Alberta and Freed, Jack H}]{Tombolato2006}
\textbf{\color{eLifeMediumGrey} Tombolato F}, Ferrarini A, Freed JH.
\newblock {Modeling the Effects of Structure and Dynamics of the Nitroxide Side
  Chain on the ESR Spectra of Spin-Labeled Proteins}.
\newblock The Journal of Physical Chemistry B.  2006 dec; 110(51):26260--26271.
\newblock \urlprefix\url{https://doi.org/10.1021/jp062949z},
  \href{10.1021/jp062949z}{\doiprefix \detokenize{10.1021/jp062949z}}.

\bibitem[{Tunyasuvunakool et~al.(2021)Tunyasuvunakool, Kathryn and Adler, Jonas
  and Wu, Zachary and Green, Tim and Zielinski, Michal and {\v{Z}}{\'{i}}dek,
  Augustin and Bridgland, Alex and Cowie, Andrew and Meyer, Clemens and Laydon,
  Agata and Velankar, Sameer and Kleywegt, Gerard J and Bateman, Alex and
  Evans, Richard and Pritzel, Alexander and Figurnov, Michael and Ronneberger,
  Olaf and Bates, Russ and Kohl, Simon A A and Potapenko, Anna and Ballard,
  Andrew J and Romera-Paredes, Bernardino and Nikolov, Stanislav and Jain,
  Rishub and Clancy, Ellen and Reiman, David and Petersen, Stig and Senior,
  Andrew W and Kavukcuoglu, Koray and Birney, Ewan and Kohli, Pushmeet and
  Jumper, John and Hassabis, Demis}]{Tunyasuvunakool2021}
\textbf{\color{eLifeMediumGrey} Tunyasuvunakool K}, Adler J, Wu Z, Green T,
  Zielinski M, {\v{Z}}{\'{i}}dek A, Bridgland A, Cowie A, Meyer C, Laydon A,
  Velankar S, Kleywegt GJ, Bateman A, Evans R, Pritzel A, Figurnov M,
  Ronneberger O, Bates R, Kohl SAA, Potapenko A, et~al.
\newblock {Highly accurate protein structure prediction for the human
  proteome}.
\newblock Nature.  2021; 596(7873):590--596.
\newblock \urlprefix\url{https://doi.org/10.1038/s41586-021-03828-1},
  \href{10.1038/s41586-021-03828-1}{\doiprefix
  \detokenize{10.1038/s41586-021-03828-1}}.

\bibitem[{Tuukkanen et~al.(2017)Tuukkanen, Anne T. and Spilotros, Alessandro
  and Svergun, Dmitri I.}]{Tuukkanen2017}
\textbf{\color{eLifeMediumGrey} Tuukkanen AT}, Spilotros A, Svergun DI.
\newblock {Progress in small-angle scattering from biological solutions at
  high-brilliance synchrotrons}.
\newblock IUCrJ.  2017; 4:518--528.
\newblock \href{10.1107/S2052252517008740}{\doiprefix
  \detokenize{10.1107/S2052252517008740}}.

\bibitem[{Unke et~al.(2021)Unke, Oliver T and Chmiela, Stefan and Sauceda,
  Huziel E and Gastegger, Michael and Poltavsky, Igor and Sch{\"{u}}tt, Kristof
  T and Tkatchenko, Alexandre and M{\"{u}}ller, Klaus-Robert}]{Unke2021}
\textbf{\color{eLifeMediumGrey} Unke OT}, Chmiela S, Sauceda HE, Gastegger M,
  Poltavsky I, Sch{\"{u}}tt KT, Tkatchenko A, M{\"{u}}ller KR.
\newblock {Machine Learning Force Fields}.
\newblock Chemical Reviews.  2021 aug; 121(16):10142--10186.
\newblock \urlprefix\url{https://doi.org/10.1021/acs.chemrev.0c01111},
  \href{10.1021/acs.chemrev.0c01111}{\doiprefix
  \detokenize{10.1021/acs.chemrev.0c01111}}.

\bibitem[{Vitalis and Pappu(2009)Vitalis, Andreas and Pappu, Rohit
  V}]{Vitalis2009}
\textbf{\color{eLifeMediumGrey} Vitalis A}, Pappu RV.
\newblock {ABSINTH: a new continuum solvation model for simulations of
  polypeptides in aqueous solutions}.
\newblock Journal of computational chemistry.  2009 apr; 30(5):673--699.
\newblock \urlprefix\url{https://pubmed.ncbi.nlm.nih.gov/18506808
  https://www.ncbi.nlm.nih.gov/pmc/articles/PMC2670230/},
  \href{10.1002/jcc.21005}{\doiprefix \detokenize{10.1002/jcc.21005}}.

\bibitem[{Wang et~al.(2014)Wang, Lee-Ping and Martinez, Todd J and Pande, Vijay
  S}]{Wang2014}
\textbf{\color{eLifeMediumGrey} Wang LP}, Martinez TJ, Pande VS.
\newblock {Building Force Fields: An Automatic, Systematic, and Reproducible
  Approach}.
\newblock The Journal of Physical Chemistry Letters.  2014 jun;
  5(11):1885--1891.
\newblock \urlprefix\url{https://doi.org/10.1021/jz500737m},
  \href{10.1021/jz500737m}{\doiprefix \detokenize{10.1021/jz500737m}}.

\bibitem[{Ward et~al.(2004)Ward, J J and Sodhi, J S and McGuffin, L J and
  Buxton, B F and Jones, D T}]{Ward2004}
\textbf{\color{eLifeMediumGrey} Ward JJ}, Sodhi JS, McGuffin LJ, Buxton BF,
  Jones DT.
\newblock {Prediction and Functional Analysis of Native Disorder in Proteins
  from the Three Kingdoms of Life}.
\newblock Journal of Molecular Biology.  2004; 337(3):635--645.
\newblock
  \urlprefix\url{https://www.sciencedirect.com/science/article/pii/S0022283604001482},
  \href{https://doi.org/10.1016/j.jmb.2004.02.002}{\doiprefix
  \detokenize{https://doi.org/10.1016/j.jmb.2004.02.002}}.

\bibitem[{Weber et~al.(2018)Weber, Benedikt and Hora, Manuel and Kazman, Pamina
  and G{\"{o}}bl, Christoph and Camilloni, Carlo and Reif, Bernd and Buchner,
  Johannes}]{Weber2018}
\textbf{\color{eLifeMediumGrey} Weber B}, Hora M, Kazman P, G{\"{o}}bl C,
  Camilloni C, Reif B, Buchner J.
\newblock {The Antibody Light-Chain Linker Regulates Domain Orientation and
  Amyloidogenicity}.
\newblock Journal of Molecular Biology.  2018; 430(24):4925--4940.
\newblock
  \urlprefix\url{https://www.sciencedirect.com/science/article/pii/S0022283618307794},
  \href{https://doi.org/10.1016/j.jmb.2018.10.024}{\doiprefix
  \detokenize{https://doi.org/10.1016/j.jmb.2018.10.024}}.

\bibitem[{Wriggers et~al.(2005)Wriggers, Willy and Chakravarty, Sugoto and
  Jennings, Patricia A}]{Wriggers2005}
\textbf{\color{eLifeMediumGrey} Wriggers W}, Chakravarty S, Jennings PA.
\newblock {Control of protein functional dynamics by peptide linkers}.
\newblock Peptide Science.  2005 jan; 80(6):736--746.
\newblock \urlprefix\url{https://doi.org/10.1002/bip.20291},
  \href{https://doi.org/10.1002/bip.20291}{\doiprefix
  \detokenize{https://doi.org/10.1002/bip.20291}}.

\bibitem[{Wright and Dyson(2015)Wright, Peter E and Dyson, H Jane}]{Wright2015}
\textbf{\color{eLifeMediumGrey} Wright PE}, Dyson HJ.
\newblock {Intrinsically disordered proteins in cellular signalling and
  regulation}.
\newblock Nature Reviews Molecular Cell Biology.  2015; 16(1):18--29.
\newblock \urlprefix\url{https://doi.org/10.1038/nrm3920},
  \href{10.1038/nrm3920}{\doiprefix \detokenize{10.1038/nrm3920}}.

\bibitem[{Wu et~al.(2018)Wu, Hao and Wolynes, Peter G and Papoian, Garegin
  A}]{Wu2018}
\textbf{\color{eLifeMediumGrey} Wu H}, Wolynes PG, Papoian GA.
\newblock {AWSEM-IDP: A Coarse-Grained Force Field for Intrinsically Disordered
  Proteins}.
\newblock The journal of physical chemistry B.  2018 dec; 122(49):11115--11125.
\newblock \urlprefix\url{https://pubmed.ncbi.nlm.nih.gov/30091924
  https://www.ncbi.nlm.nih.gov/pmc/articles/PMC6713210/},
  \href{10.1021/acs.jpcb.8b05791}{\doiprefix
  \detokenize{10.1021/acs.jpcb.8b05791}}.

\bibitem[{Xue et~al.(2012)Xue, Bin and Dunker, A. Keith and Uversky, Vladimir
  N.}]{Xue2012}
\textbf{\color{eLifeMediumGrey} Xue B}, Dunker AK, Uversky VN.
\newblock {Orderly order in protein intrinsic disorder distribution: Disorder
  in 3500 proteomes from viruses and the three domains of life}.
\newblock Journal of Biomolecular Structure and Dynamics.  2012;
  30(2):137--149.
\newblock \href{10.1080/07391102.2012.675145}{\doiprefix
  \detokenize{10.1080/07391102.2012.675145}}.

\bibitem[{Yang et~al.(2021)Yang, Huan and Xiong, Zhaoping and Zonta,
  Francesco}]{Yang2021.04.26.441401}
\textbf{\color{eLifeMediumGrey} Yang H}, Xiong Z, Zonta F.
\newblock {Construction of a neural network energy function for protein
  physics}.
\newblock bioRxiv.  2021;
  \urlprefix\url{https://www.biorxiv.org/content/early/2021/04/27/2021.04.26.441401},
  \href{10.1101/2021.04.26.441401}{\doiprefix
  \detokenize{10.1101/2021.04.26.441401}}.

\bibitem[{Yang et~al.(2010)Yang, Sichun and Blachowicz, Lydia and Makowski, Lee
  and Roux, Beno{\^{i}}t}]{Yang2010}
\textbf{\color{eLifeMediumGrey} Yang S}, Blachowicz L, Makowski L, Roux B.
\newblock {Multidomain assembled states of Hck tyrosine kinase in solution}.
\newblock Proceedings of the National Academy of Sciences.  2010 aug;
  \urlprefix\url{http://www.pnas.org/content/early/2010/08/18/1004569107.abstract},
  \href{10.1073/pnas.1004569107}{\doiprefix
  \detokenize{10.1073/pnas.1004569107}}.

\bibitem[{Zerze et~al.(2019)Zerze, G{\"{u}}l H and Zheng, Wenwei and Best,
  Robert B and Mittal, Jeetain}]{Zerze2019}
\textbf{\color{eLifeMediumGrey} Zerze GH}, Zheng W, Best RB, Mittal J.
\newblock {Evolution of All-Atom Protein Force Fields to Improve Local and
  Global Properties}.
\newblock The Journal of Physical Chemistry Letters.  2019 may;
  10(9):2227--2234.
\newblock \urlprefix\url{https://doi.org/10.1021/acs.jpclett.9b00850},
  \href{10.1021/acs.jpclett.9b00850}{\doiprefix
  \detokenize{10.1021/acs.jpclett.9b00850}}.

\bibitem[{Zweckstetter(2008)Zweckstetter, Markus}]{Zweckstetter2008}
\textbf{\color{eLifeMediumGrey} Zweckstetter M}.
\newblock {NMR: prediction of molecular alignment from structure using the
  PALES software}.
\newblock Nature Protocols.  2008; 3(4):679--690.
\newblock \urlprefix\url{https://doi.org/10.1038/nprot.2008.36},
  \href{10.1038/nprot.2008.36}{\doiprefix \detokenize{10.1038/nprot.2008.36}}.

\end{thebibliography}

\end{document}